\documentclass{aa}
\usepackage{graphics,color}
\usepackage{epsf,epsfig}
\usepackage{rotating}
\usepackage{rotate}
\begin{document}
%
%
\title{
Bispectrum speckle interferometry of \object{IRC\,+10\,216}: \\
the dynamic evolution of the innermost circumstellar environment
from 1995 to 2001
}
\author{
G. Weigelt\inst{1}\and
Y.Y.\ Balega\inst{2}\and
T.\ Bl\"ocker\inst{1}\and
K.-H.\ Hofmann\inst{1}\and
A.B.\ Men'shchikov\inst{1}\and
J.M.\ Winters\inst{1}
}
\institute{
Max-Planck-Institut f\"ur Radioastronomie, Auf dem H\"ugel 69,
D--53121 Bonn, Germany \\
(email: {\tt bloecker@mpifr-bonn.mpg.de,
 hofmann@mpifr-bonn.mpg.de, sasha@mpifr-bonn.mpg.de,} \\
 {\tt jwinters@mpifr-bonn.mpg.de})
\and
Special Astrophysical Observatory, Nizhnij Arkhyz, Zelenchuk region,
Karachai--Cherkesia, 35147, Russia \\ (email: {\tt balega@sao.ru})
}
\offprints{G.\ Weigelt, \protect{\\} email: {\tt weigelt@mpifr-bonn.mpg.de}}
%

\date{Received date /  accepted date}
\titlerunning{Bispectrum speckle interferometry
 of \object{IRC\,+10\,216}
}
\authorrunning{G. Weigelt et al.}
\abstract{
We present new
near-infrared ($JHK$) bispectrum speckle-interferometry monitoring
of the carbon star \object{IRC+10216} obtained between 1999 and 2001
with the SAO 6\,m telescope.
The  $J$-, $H$-, and $K$-band resolutions are
50\,mas, 56\,mas, and 73\,mas, respectively.
The total sequence of $K$-band observations covers now 8 epochs
from 1995 to 2001
and shows the dynamic evolution of the inner dust shell.
The present observations show that the appearance of the dust shell
has considerably changed compared to the epochs of 1995 to 1998.
Four main components within a 0\farcs2 radius can be identified
in the $K$-band images.
The apparent separation of the two
initially brightest components A and B increased
from $\sim 191$ mas in 1995 to $\sim 351$ mas in 2001.
Simultaneously, component B has been fading and almost disappeared in 2000
whereas the initially faint components C and D became brighter (relative to
peak intensity).
The changes of the images can be related to changes of the optical depth
caused, for instance, by mass-loss variations or new dust condensation in the
wind. Our recent two-dimensional radiative transfer model of
\object{IRC\,+10\,216} suggests that the observed relative motion of
components A and B is not consistent with the outflow of gas and dust at the
well-known terminal wind velocity of 15 km\,s$^{-1}$. The apparent motion with
a deprojected velocity of 19 km\,s$^{-1}$ on average and of recently
27\,km\,s$^{-1}$ appears to be
caused by a displacement of the dust density peak due to dust
evaporation in the optically thicker and hotter environment.
The present monitoring, covering more than  3 pulsation
periods, shows that the structural variations are not related to the
stellar pulsation cycle in a simple way.
This is consistent with the predictions of
hydrodynamical models that enhanced dust formation takes place on a timescale
of several pulsation periods. The timescale of the fading of
component B can well be explained by the formation of new dust in the
circumstellar envelope.
}
\maketitle
\section{Introduction}  \label{Sintro}
The carbon star \object{IRC\,+10216} is a long-period variable star
evolving along the Asymptotic Giant Branch (AGB).
It is the nearest carbon star known
(distance $\sim 110-150$\,pc; Crosas \& Menten \cite{CroMen97},
Groenewegen et al.\ \cite{GroeEtal98}) and
the brightest 12\,$\mu$m object outside the solar system (IRAS 1986).
A strong stellar wind has led to an  almost complete obscuration of the
star by dust. The mass-loss rate
as measured in CO rotational lines amounts to
$2-5 \times 10^{-5} M_{\odot}$/yr (Loup et al. \cite{LoupEtal93}).
Detailed two-dimensional (2D) radiative transfer modeling shows
that \object{IRC\,+10216} had recently suffered
from an even higher mass-loss rate of
$\sim 10^{-4} M_{\odot}$/yr (Men'shchikov et al. \cite{mbbow2001}).
Based on the high mass-loss rate, long period of $P=649$\,d
(Le Bertre \cite{LeB92}), and carbon-rich  chemistry of the dust-shell,
\object{IRC\,+10\,216} is obviously in a very advanced stage of its
AGB evolution (see, e.g., Bl\"ocker \cite{Bloe99}).

\begin{figure*}
\begin{center}
\epsfxsize=44mm
\mbox{\epsffile{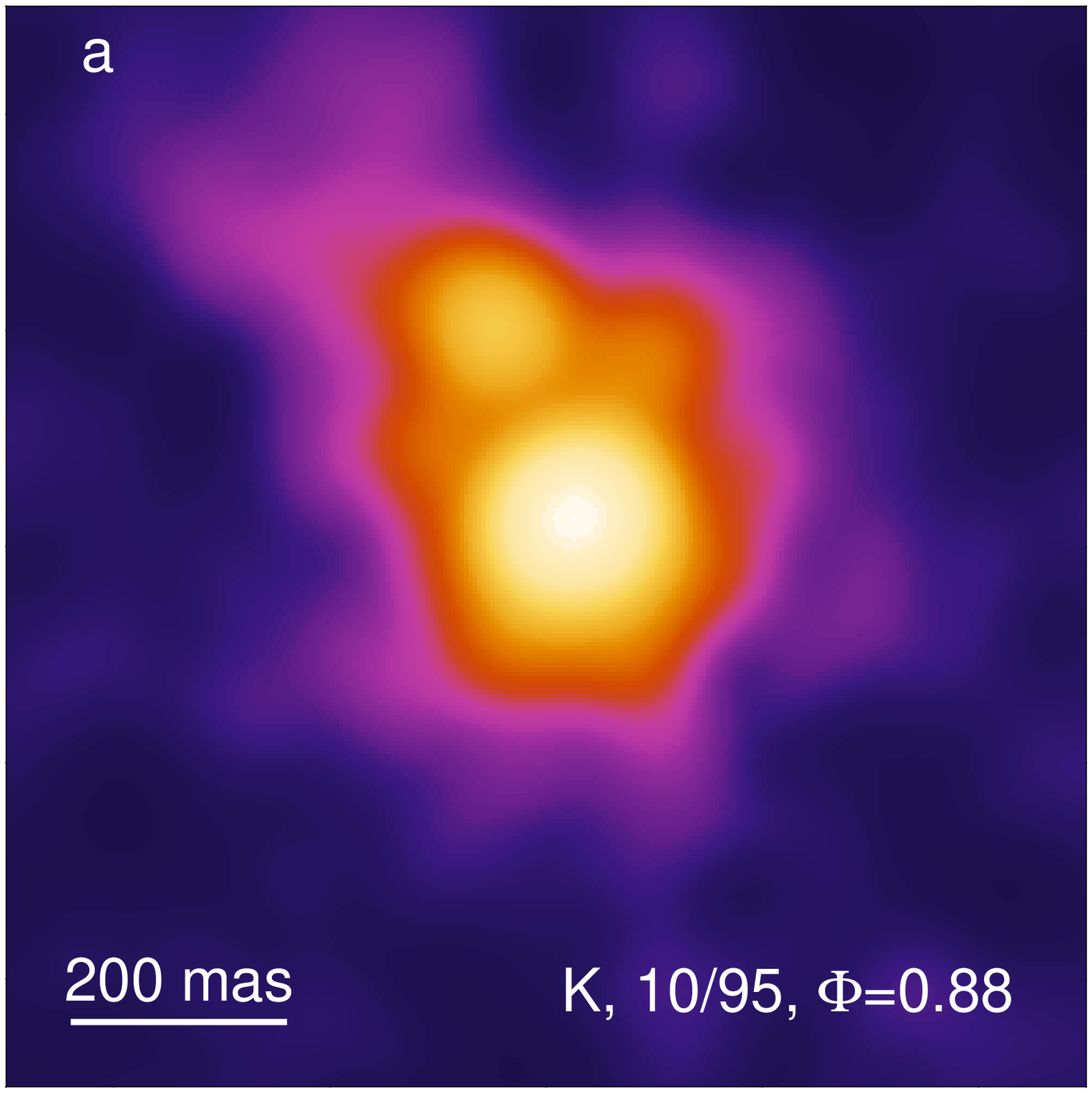}}
\epsfxsize=44mm
\mbox{\epsffile{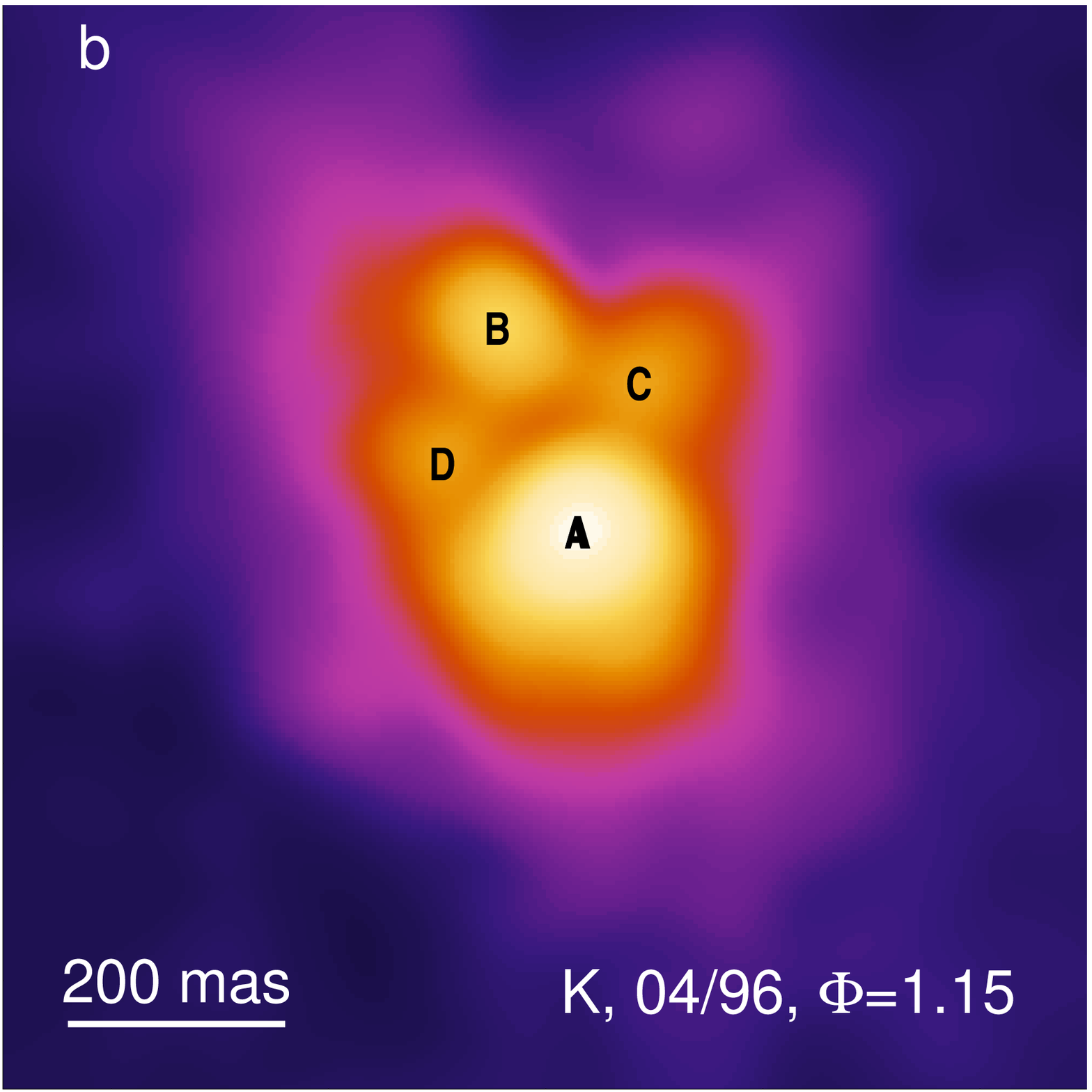}}
\epsfxsize=44mm
\mbox{\epsffile{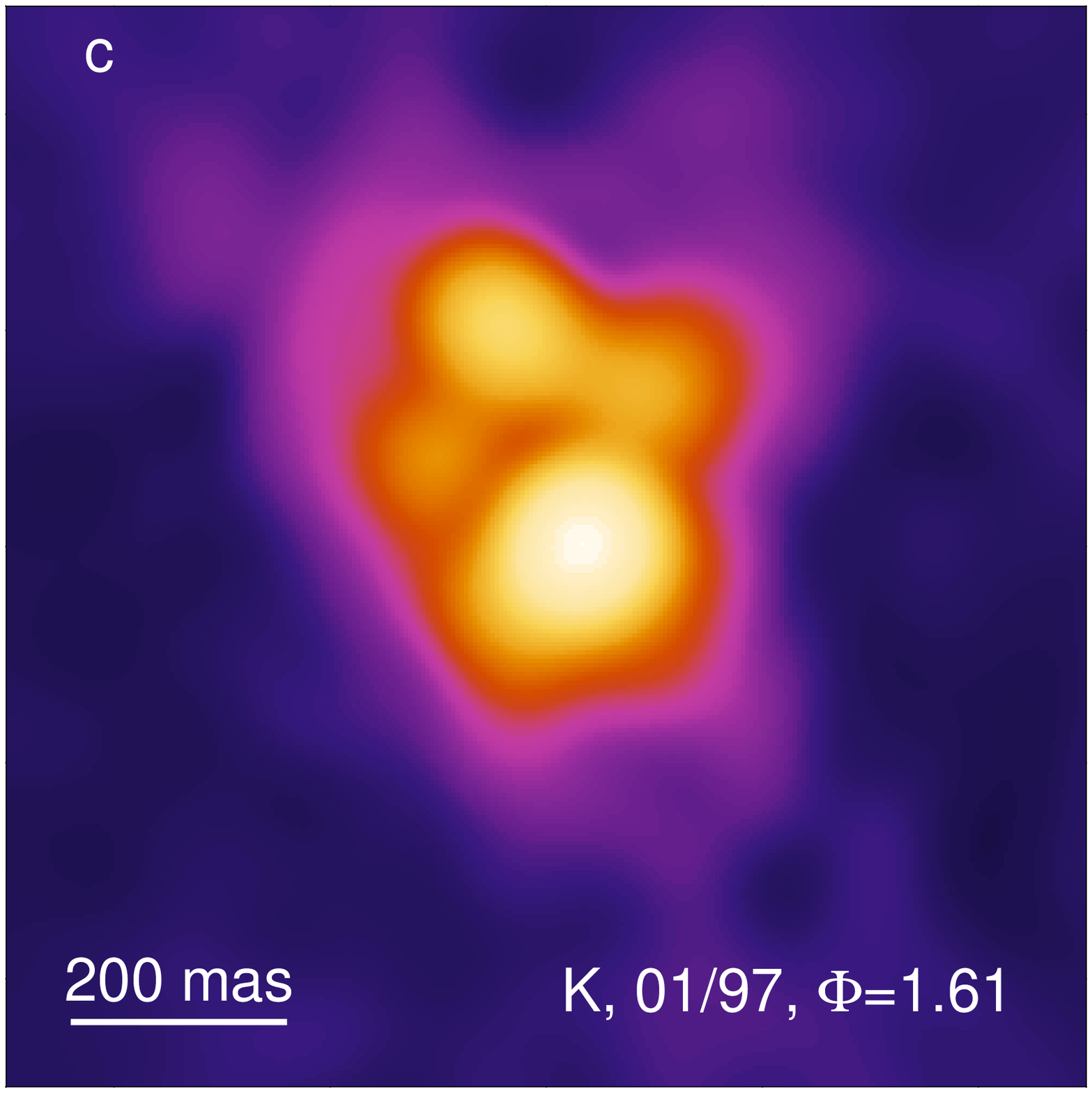}}
\epsfxsize=44mm
\mbox{\epsffile{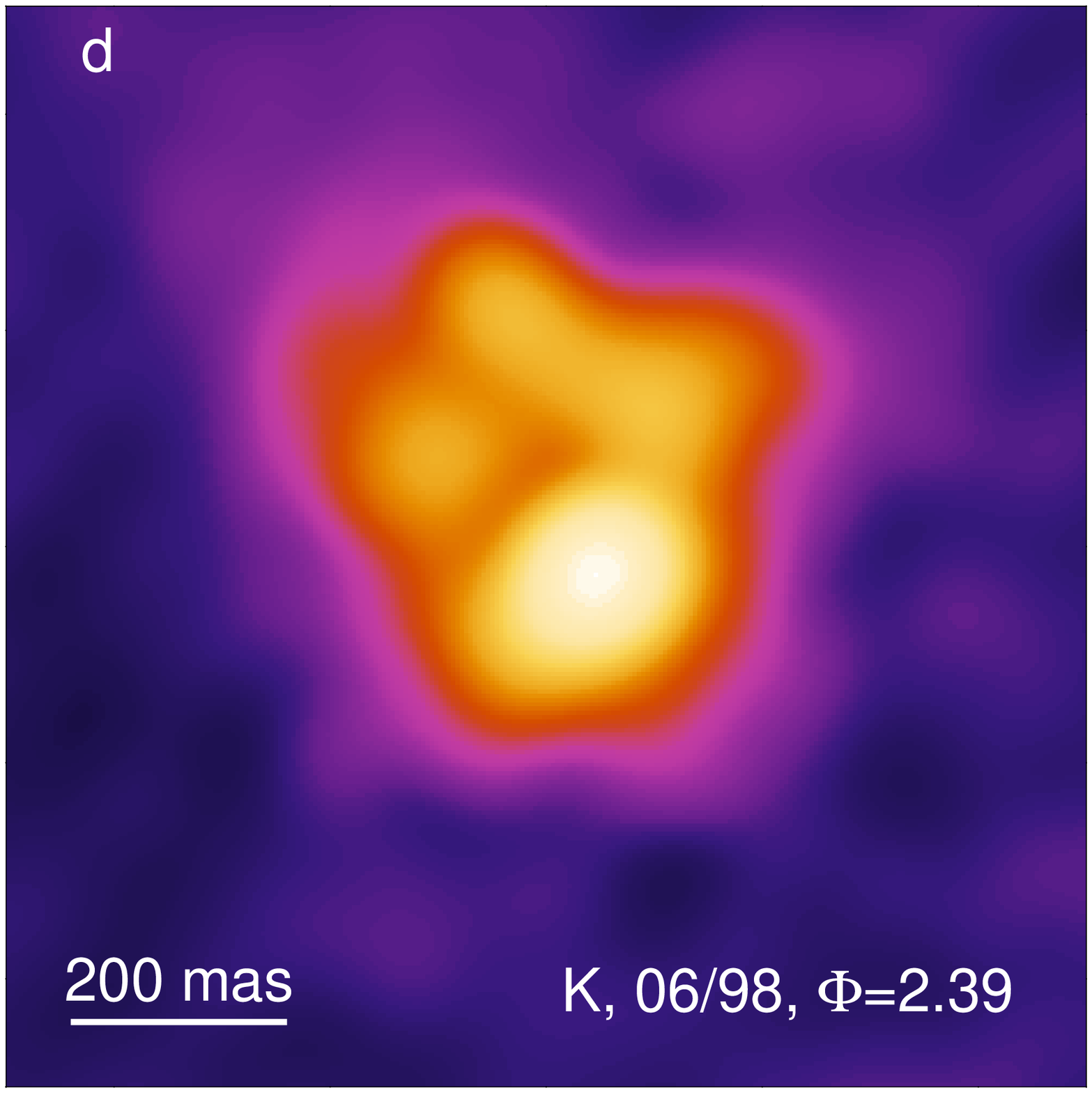}}\\[0.3mm]
\epsfxsize=44mm
\mbox{\epsffile{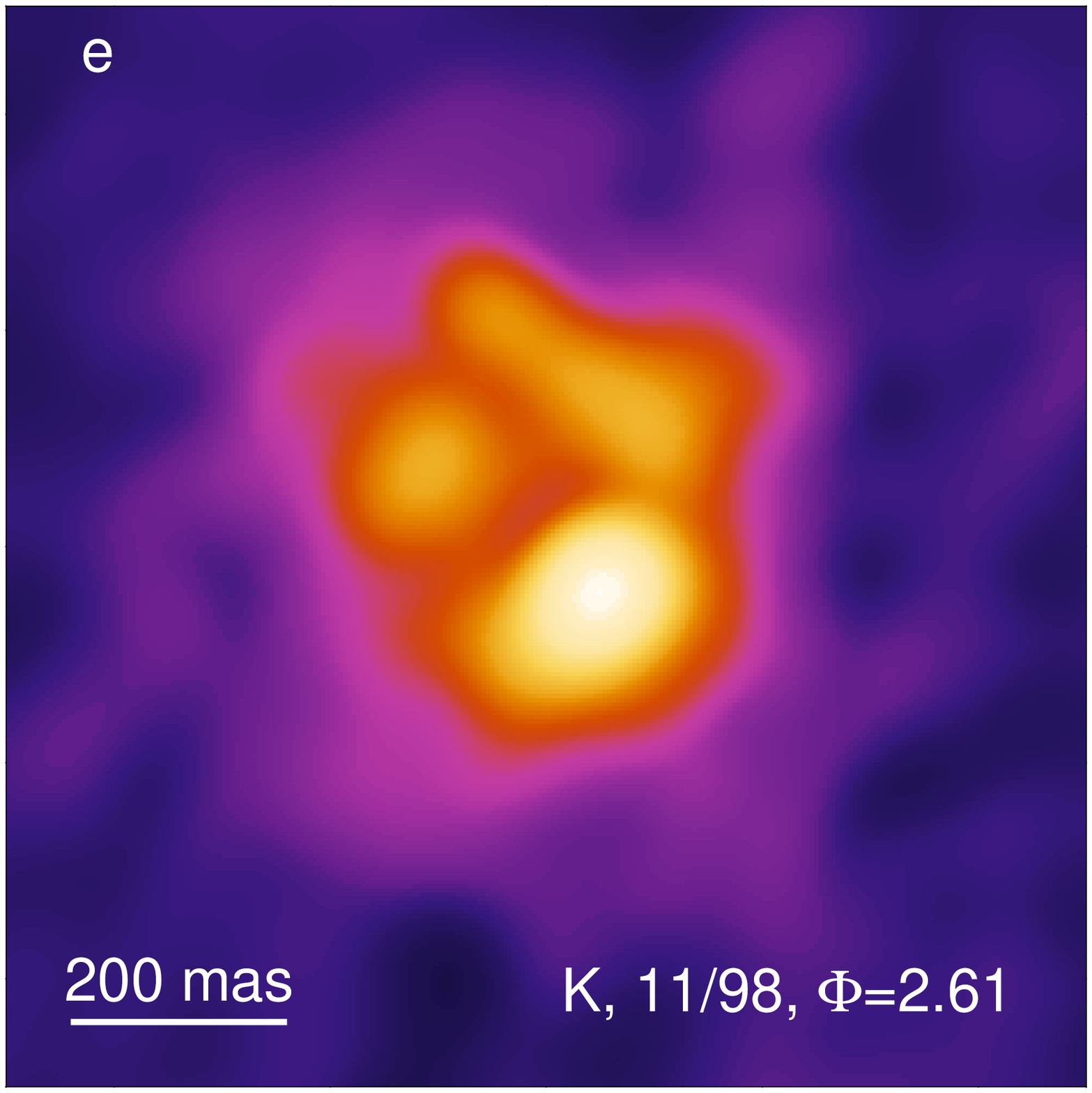}}
\epsfxsize=44mm
\mbox{\epsffile{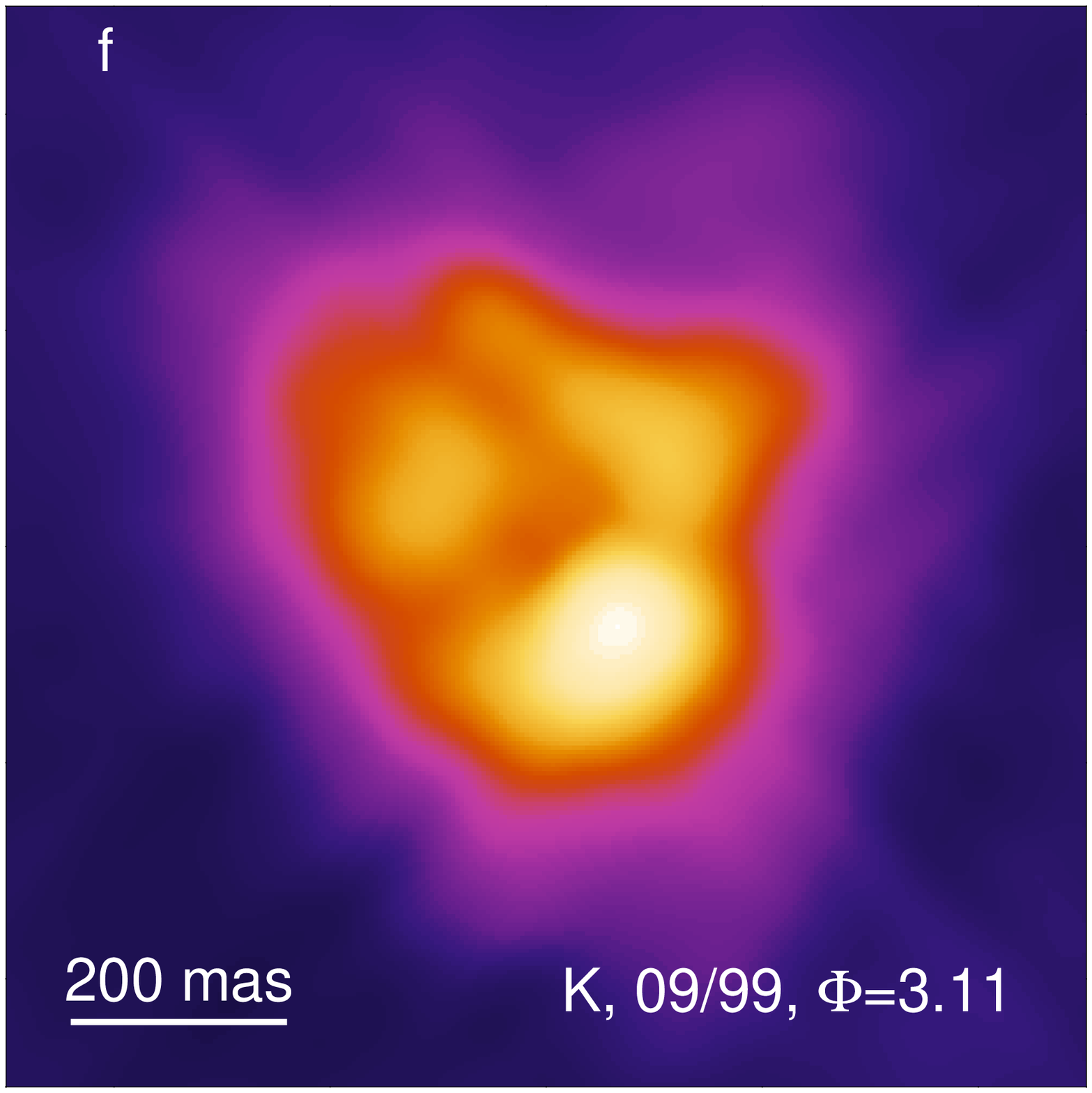}}
\epsfxsize=44mm
\mbox{\epsffile{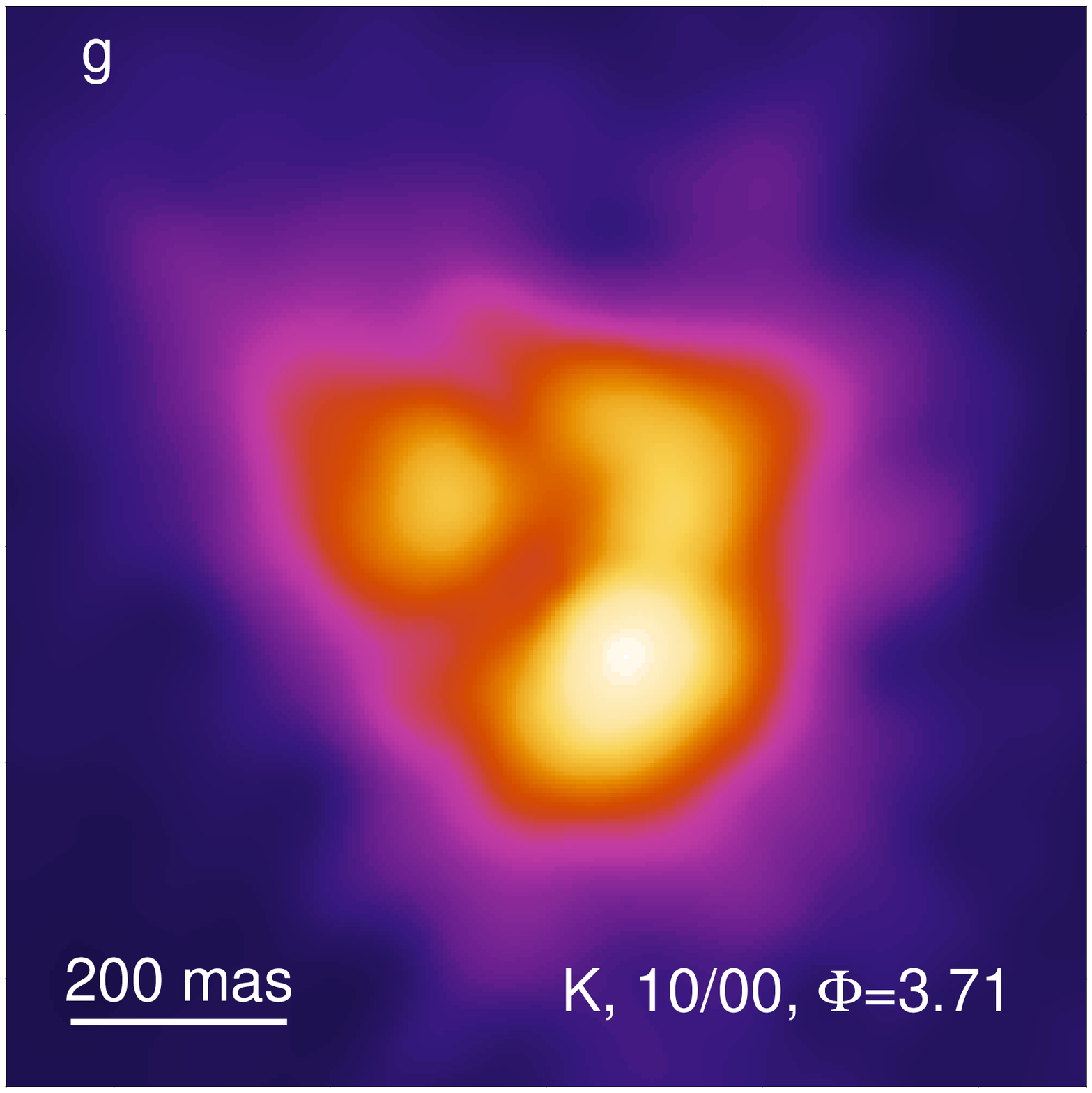}}
\epsfxsize=44mm
\mbox{\epsffile{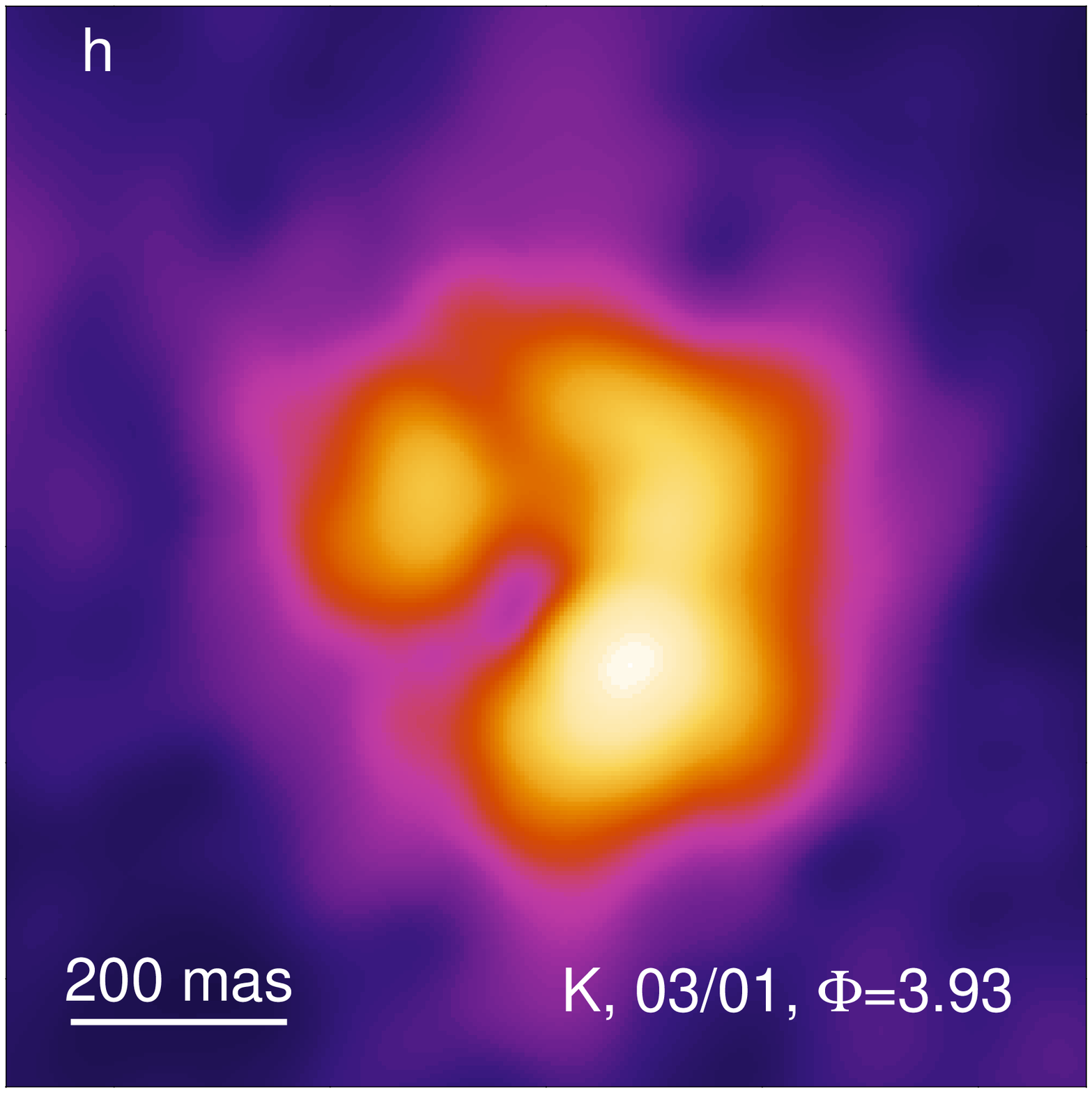}}
\end{center}
\caption{
$K$-band speckle reconstructions of \object{IRC\,+10\,216}
for 8 epochs from 1995 to 2001.
The total area is 1\arcsec$\times$1\arcsec.
All images are normalized to the brightest pixel and are
presented with the same color table.
North is up and east is to the left.
\vspace*{0.3cm}
}
\label{FKima}
%
\begin{center}
\epsfxsize=59mm
\mbox{\epsffile{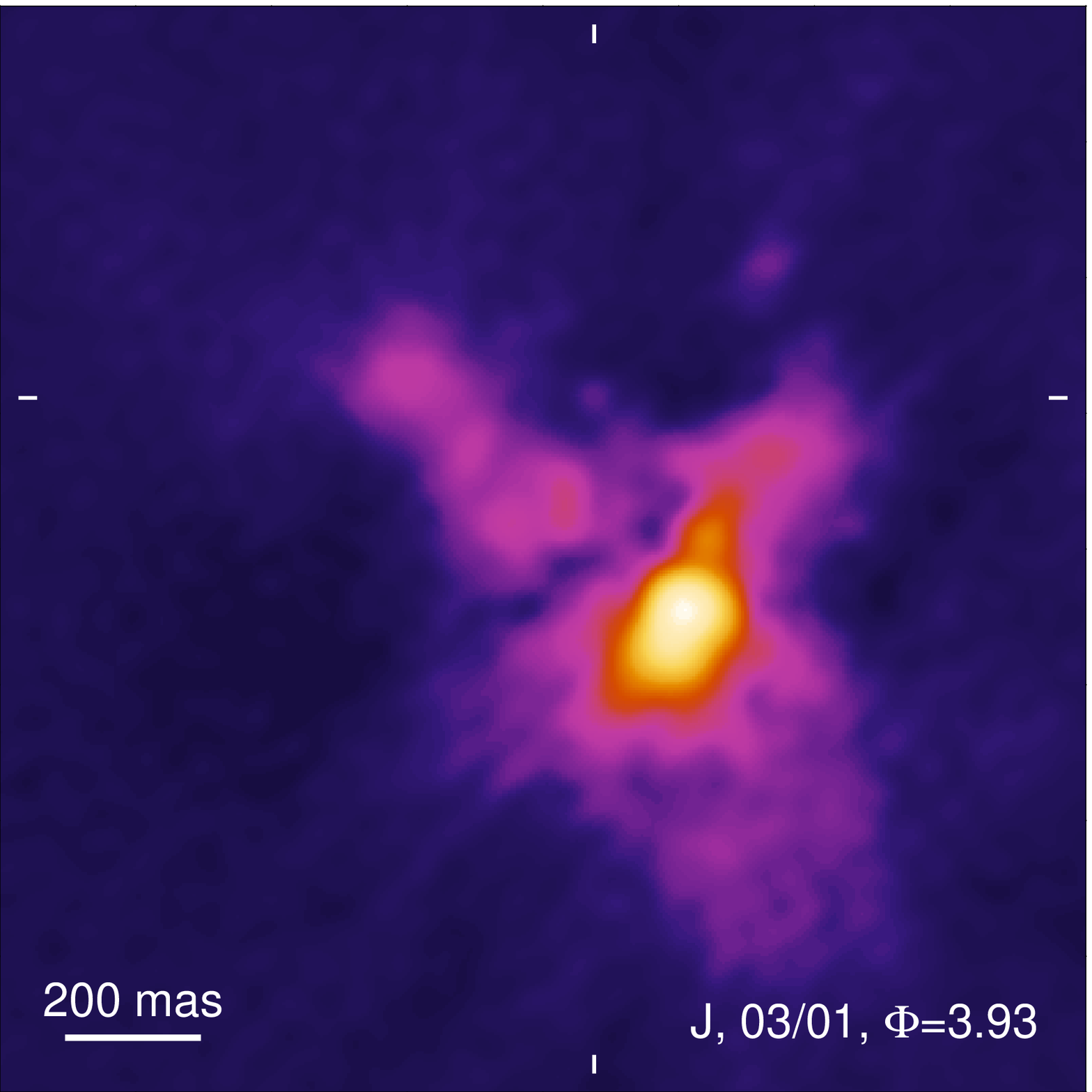}}
\epsfxsize=59mm
\mbox{\epsffile{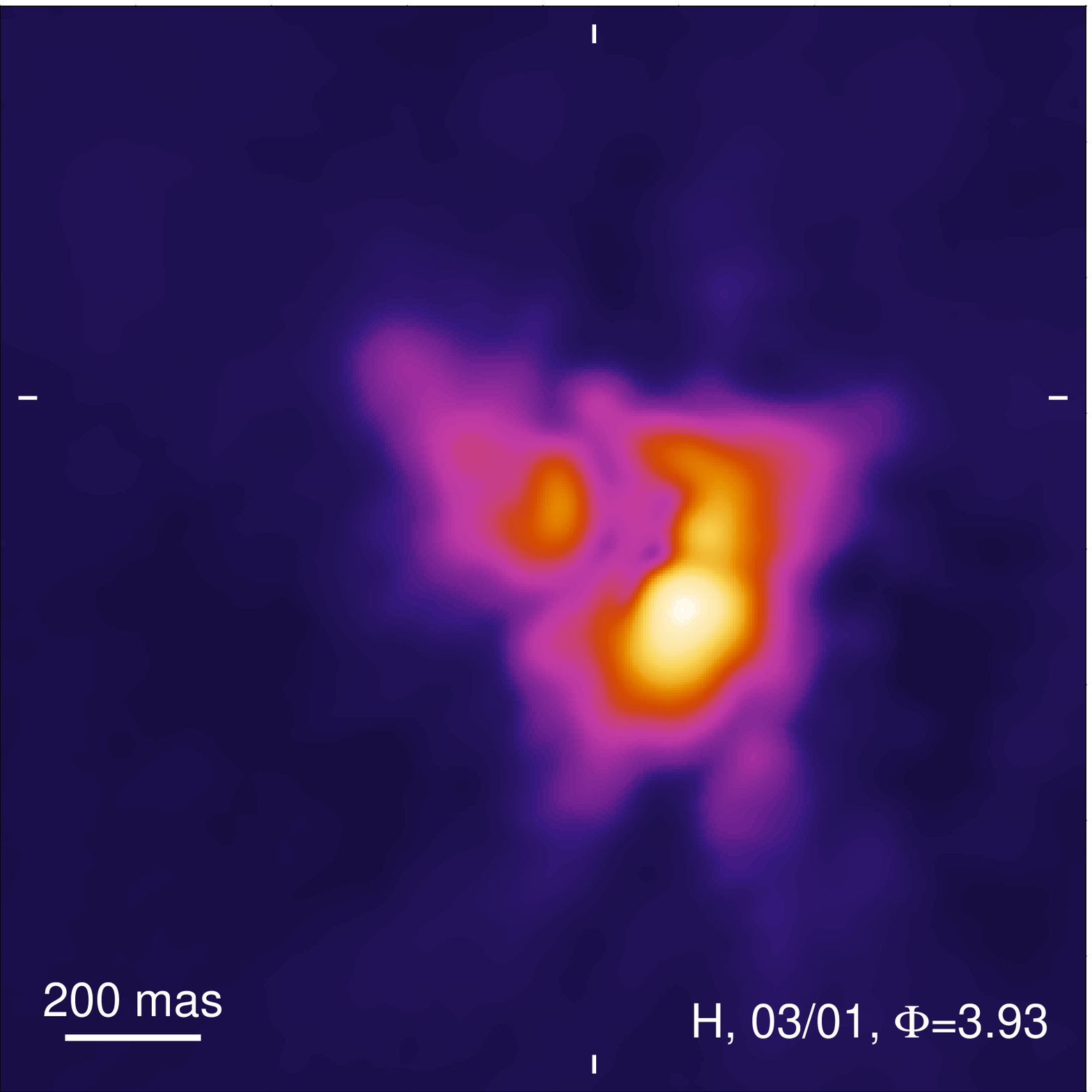}}
\epsfxsize=59mm
\mbox{\epsffile{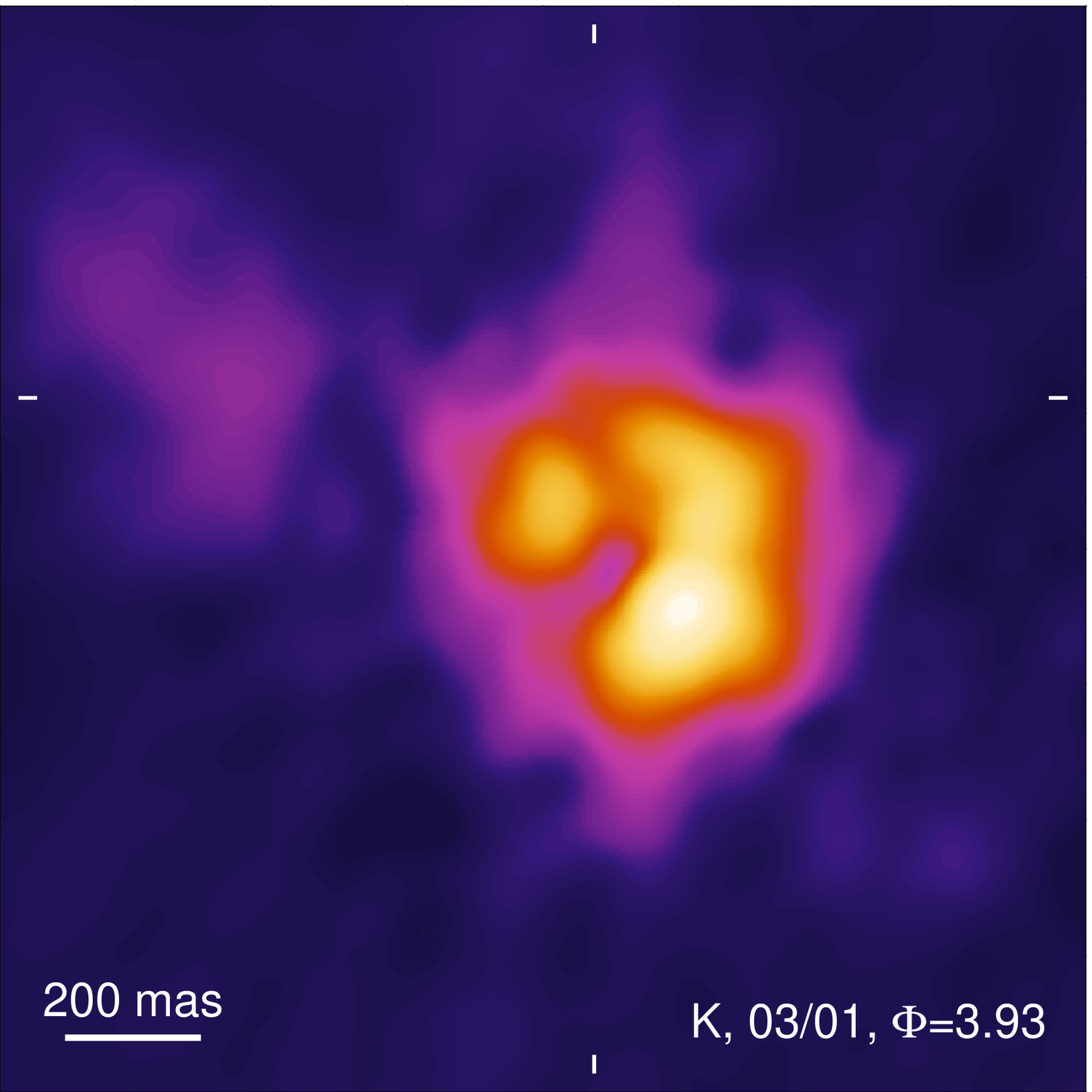}} \\[0.3mm]
\end{center}
\caption{
$J$-, $H$- and $K$-band speckle reconstructions of
\object{IRC\,+10\,216} in March 2001.
The total area is 1.6\arcsec$\times$1.6\arcsec.
All images are normalized to the brightest pixel and are
presented with the same color table.
North is up and east is to the left.
The tick marks indicate the likely position of the
central star.
}
\label{FJHKima}
\end{figure*}

%
\begin{table*}
\caption{Observational parameters. JD refers to the Julian date and
$\Phi$ to the photometric phase.
Phase 0 (maximum light)
corresponds to JD=2449430 (Mar 18, 1994) as extrapolated
from the light curve of Le Bertre (\cite{LeB92}) with $P=649$\,d.
$\lambda_\mathrm{c}$ is the central wavelength and  $\Delta\lambda$
the FWHM bandwidth of the filters.
$N_\mathrm{T}$ and $N_\mathrm{R}$
are the numbers of \object{IRC +10 216} speckle interferograms
and reference-star speckle interferograms, respectively.
$T$ is the exposure time per frame,
$S$ is the seeing (FWHM), $p$ is the pixel size, and $R$ is the resolution.
In the last column, the reference stars are given.
The observations of 1995--1998 were already presented in
Paper II 
and are given here for completeness.
} \label{obstab}
\begin{center}
\begin{tabular}{ 
                l
                c
                c
                c
                c
                c
                r
                r
                r
                c
                c
                c
                l
               }
\hline
 & Date & JD & $\Phi$ & $\lambda_\mathrm{c}$ & $\Delta\lambda$ &
$N_{\rm T}$ & $N_{\rm R}$ & $T$\hspace*{1.2mm}& $S$ & $p$ & $R$ & Ref.\ star\\
 &      &    &      & [$\mu$m]             &  [$\mu$m]         &
            &             & [ms]  & [$^{\prime\prime}$]  & [mas] & [mas] & \\
\noalign{\smallskip}
\hline\noalign{\smallskip}
$J$ & ~2 Apr 1996 & 2450176 & 1.15 & 1.24 & 0.28 & 1196 &  981 & 200  & 1.2 & 14.6 & 149 & \object{HIP 51133} \\
    & 10 Mar 2001 & 2451979 & 3.93 & 1.24 & 0.14 & 1042 &  783 & 160  & 1.0 & 13.3 & 50  & \object{HD 83871} \\
\noalign{\smallskip}\hline\noalign{\smallskip}
$H$ & 23 Jan 1997 & 2450472 & 1.61 & 1.64 & 0.31 & 1665 & 2110 & 100  & 1.5 & 19.8 & 70  & \object{HIP 52689} \\
    & 10 Mar 2001 & 2451979 & 3.93 & 1.65 & 0.32 &  607 &  915 &  30  & 1.0 & 20.1 & 56  & \object{HD 83871} \\
\noalign{\smallskip}\hline\noalign{\smallskip}
$K$ & ~8 Oct 1995 & 2449999 & 0.88 & 2.12 & 0.02 &  251 &  266 & 100  & 1.5 & 31.5 & 92  & \object{SAO 116569} \\
    & ~3 Apr 1996 & 2450177 & 1.15 & 2.17 & 0.33 & 1403 & 1363 &  70  & 2.5 & 14.6 & 82  & \object{HIP 51133} \\
    & 23 Jan 1997 & 2450472 & 1.61 & 2.19 & 0.41 & 2165 & 1539 &  50  & 0.9 & 30.6 & 87  & \object{HIP 52689} \\
    & 14 Jun 1998 & 2450979 & 2.39 & 2.17 & 0.33 &  800 &  571 &  50  & 1.6 & 30.6 & 87  & \object{HIP 50792} \\
    & ~3 Nov 1998 & 2451121 & 2.61 & 2.20 & 0.20 & 1087 &  842 &  40  & 1.3 & 27.2 & 75  & \object{HIP 49583} \\
    & 24 Sep 1999 & 2451446 & 3.11 & 2.12 & 0.21 & 2702 & 1383 &  80  & 0.9 & 26.4 & 73  & \object{HIP 49637} \\
    & 15 Oct 2000 & 2451833 & 3.70 & 2.09 & 0.02 & 1740 & 2091 &  30  & 1.3 & 26.8 & 73  & \object{HIP 49637}\\
    & ~9 Mar 2001 & 2451978 & 3.93 & 2.09 & 0.02 &  390 &  777 &  20  & 1.0 & 27.0 & 73  & \object{31 Leo} \\
\hline
\end{tabular}
\end{center}
\end{table*}
Interferometric near-infrared imaging
of \object{IRC+10216} with angular resolutions of better than 100 mas
has revealed that its dust shell is clumpy
and bipolar, and is changing on a time scale of only $\sim$1\,yr
(Weigelt et al.\ \cite{WeiEtal97},
 Weigelt et al.\ \cite{WeiEtal98} [hereafter Paper~I],
 Haniff \& Buscher \cite{HanBus98},
 Osterbart et al.\ \cite{OstEtal00} [hereafter Paper~II],
 Tuthill et al.\ \cite{TutEtal00}).
In 1996, four components were identified in the inner dust shell of
\object{IRC\,+10\,216} within a radius of 200\,mas
(Weigelt et al. \cite{WeiEtal97}, Paper~I, Haniff \& Buscher
\cite{HanBus98}) and were denoted as A, B, C, and D
in order of decreasing brightness (see Fig.~\ref{FKima}).
On larger scales the envelope of \object{IRC\,+10216}
appears to be spherically symmetric
(Mauron \& Huggins \cite{MauHug99,MauHug00}).
Since most dust shells around AGB stars are known to be spherically
symmetric on larger scales,
whereas most proto-planetary nebulae (PPN) appear in
axisymmetric geometry (Olofsson \cite{Olof96}),
it is likely that \object{IRC\,+10216}
has already entered the transition phase to PPN. This suggests
that the break of symmetry already takes place at the end of
the AGB evolution.
So far, only a few AGB objects are known to show prominent
asphericities of their dust shells in the near-infrared,
and are therefore believed to have
entered this transition phase at the end of their AGB life. This
includes, for instance, the carbon star \object{CIT\,6}
(Monnier et al.\ \cite{MonEtal00}),
and the oxygen-rich stars \object{AFGL\,2290} (Gauger et al.\ \cite{GauEtal99})
and \object{CIT\,3} (Hofmann et al.\ \cite{HofEtal01}).

%

Recently, we have developed a 2D radiative transfer model of \object{IRC\,+10\,216}
(Men'shchikov et al. \cite{mbbow2001}; hereafter Paper III) which can explain
many aspects of the nebula. In this model, the star is located at component B
and is surrounded by an optically thick shell with bipolar cavities. The
brightest southern component A is identified with radiation emitted and
scattered in the optically thinner southern cavity of the dense shell. However,
Weigelt et al. (\cite{WeiEtal98}), Haniff \& Buscher (\cite{HanBus98}), and
Tuthill et al. (\cite{TutEtal00}) argued that the star is located at the
position of component A and that components B, C, and D are dust clouds. In the
present paper, we attempt to explain the recent changes in the nebula in light
of the 2D model of Paper III, and will comment on the alternative
interpretations only in passing.
%

High-resolution near-infrared monitoring of the components A, B, C, and D
has already revealed that the dust shell of \object{IRC\,+10\,216}
is rapidly evolving. The 6\,m telescope
speckle observations presented in Paper~II
cover five epochs between 1995 and 1998 and
show that the separations between the different components had steadily been
increasing. For example, the distance between
the initially brightest components A and B increased by 36\% during 1995-1998.
These results are in very good agreement with
Keck telescope $K$-band observations obtained
with the highest resolution to date
at 7 epochs between
1997 and 1999
by Tuthill et al.\ (\cite{TutEtal00}).
Such direct observations of the dust-shell evolution
offer an ideal opportunity to study
the mass-loss process in the late stages of AGB evolution,
revealing details of the dust formation process
as well as the geometry and clumpiness of the
stellar wind.

This paper presents new near-infrared bispectrum speckle interferometry
monitoring
of \object{IRC+10216} obtained with the SAO 6\,m telescope in 1999--2001,
adding 3 new epochs to the already available data of dust-shell evolution.
The present observations show that the appearance of the dust shell
has considerably changed compared to the epochs of 1995 to 1998.
This paper is organized as follows. In Sect.~\ref{Sobs},
$K$-band observations
of the dust-shell evolution from 1999 to 2001,
a comparison of  $J$-, $H$-, and $K$-band images of  1996/97 and 2001,
and $J{-}H$, $J{-}K$, and $H{-}K$ color images are presented.
In Sect.~\ref{Sdis}, these observations are discussed on the basis of
the general morphology,                                  
2D radiative transfer models,            
and dust-formation models.             
Conclusions are given in Sect.~\ref{Scon}.
\section{Observations} \label{Sobs}
\subsection{Data reduction}
\begin{figure*}
\begin{center}
\epsfxsize=55mm
\mbox{\epsffile{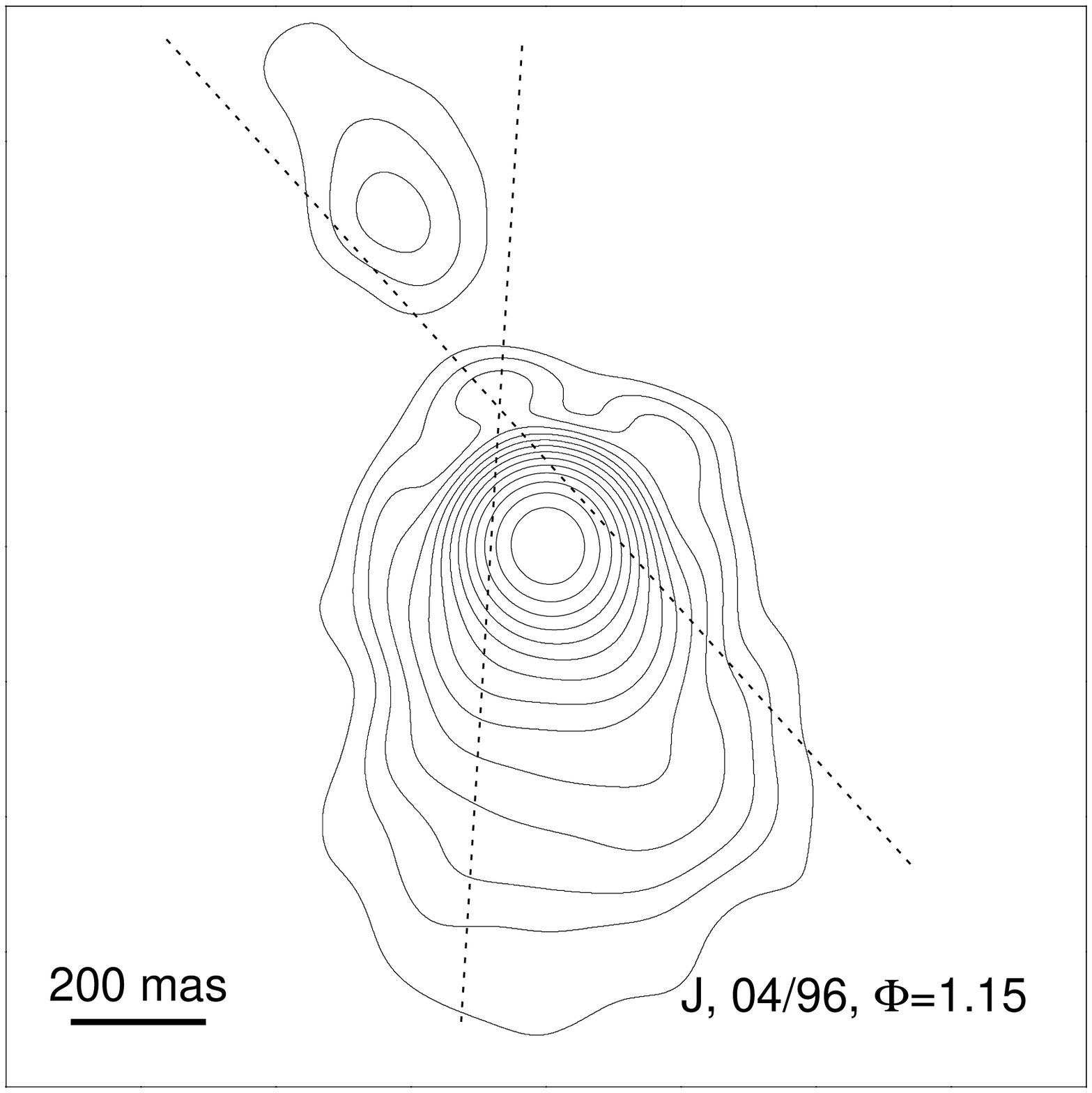}}
\epsfxsize=55mm
\mbox{\epsffile{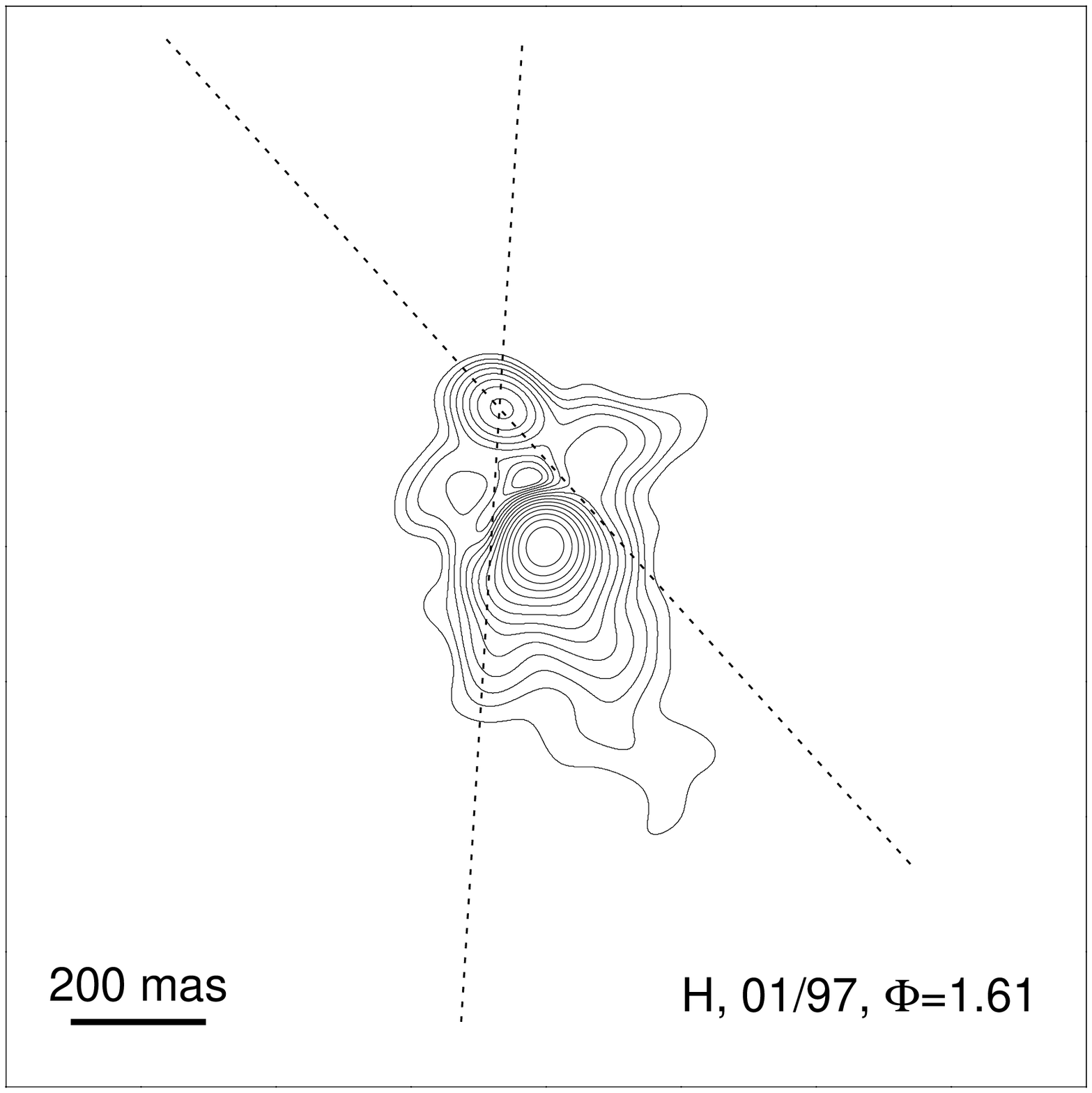}}
\epsfxsize=55mm
\mbox{\epsffile{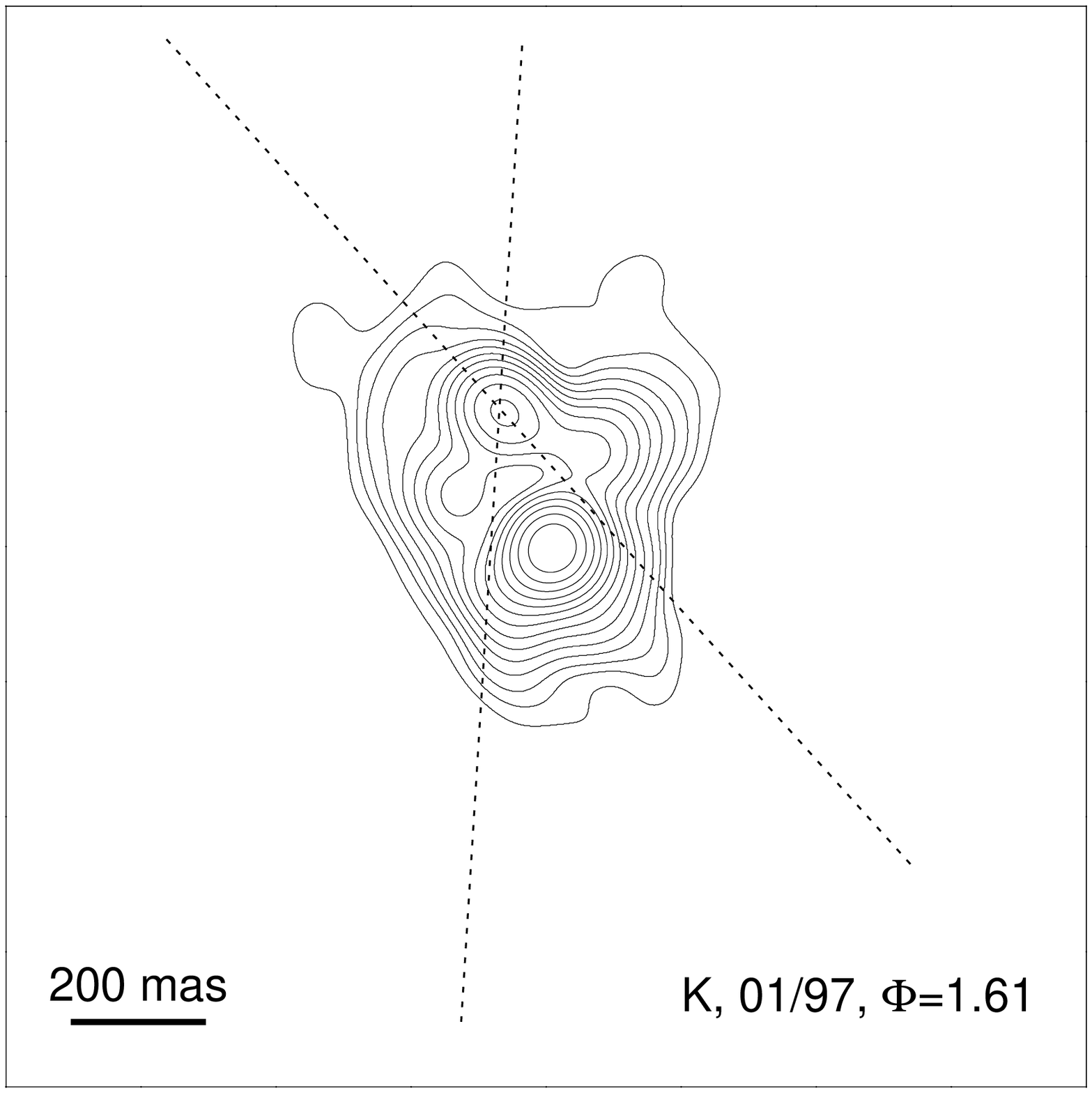}} \\[0.3mm]
\epsfxsize=55mm
\mbox{\epsffile{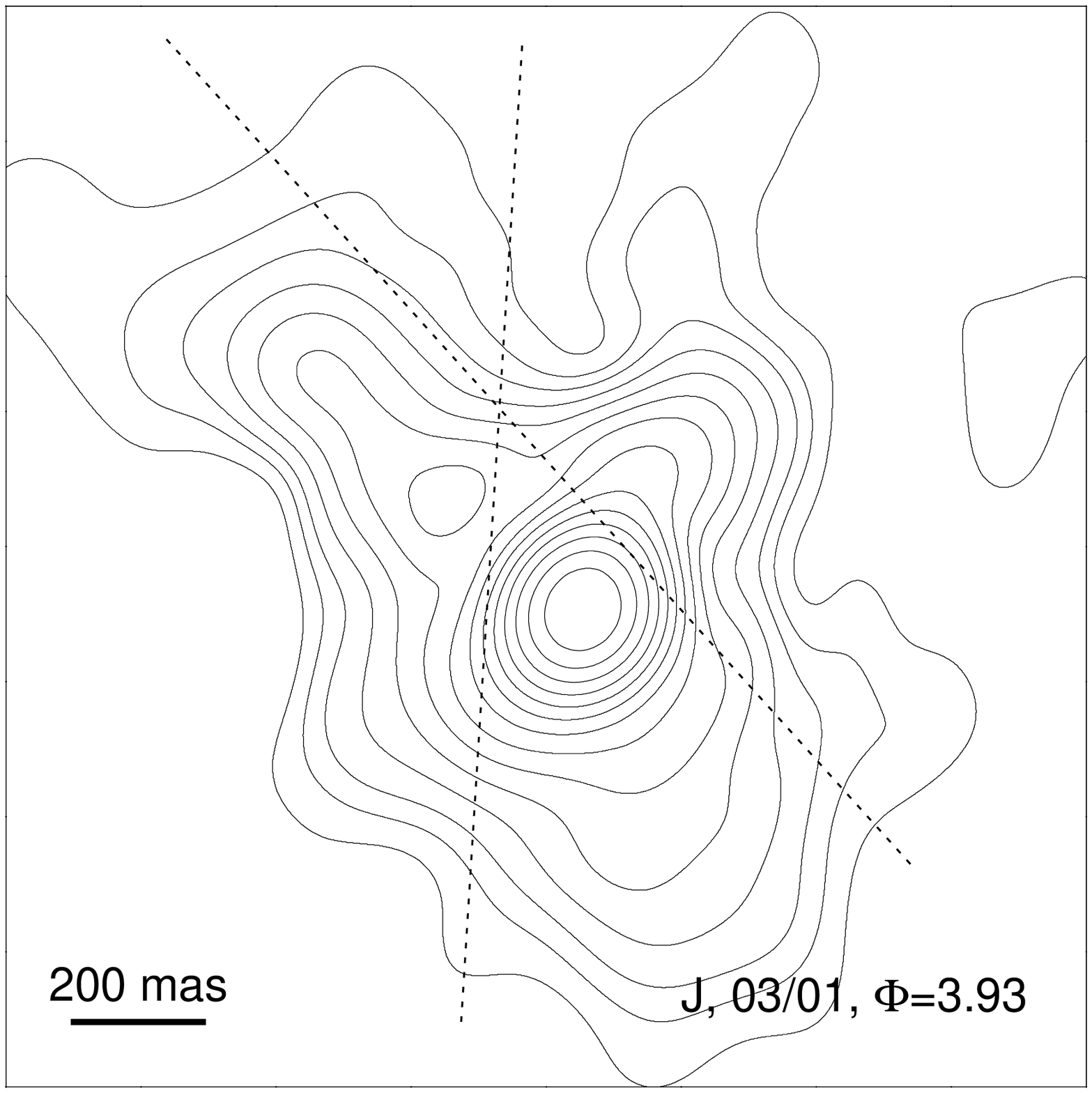}}
\epsfxsize=55mm
\mbox{\epsffile{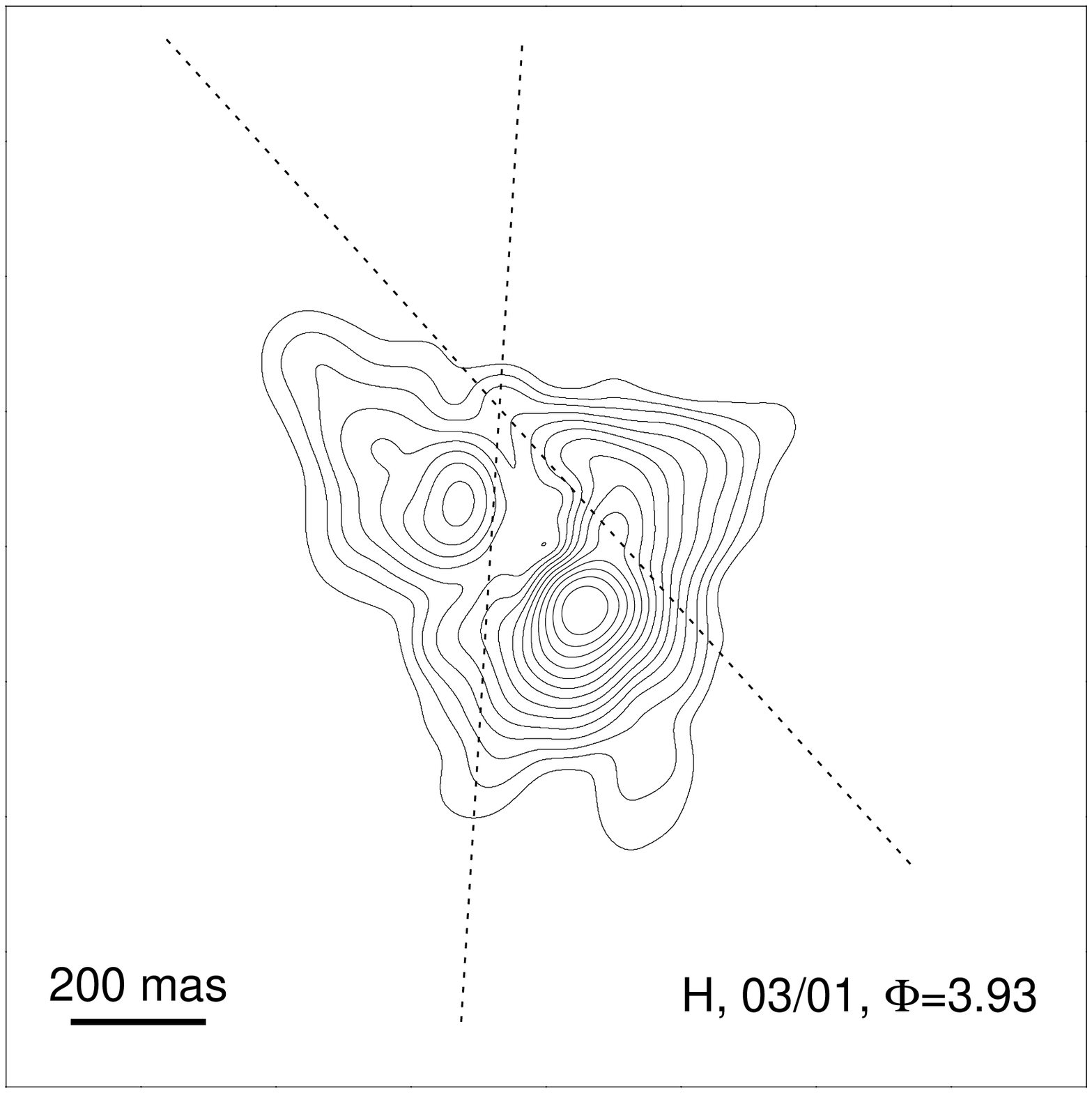}}
\epsfxsize=55mm
\mbox{\epsffile{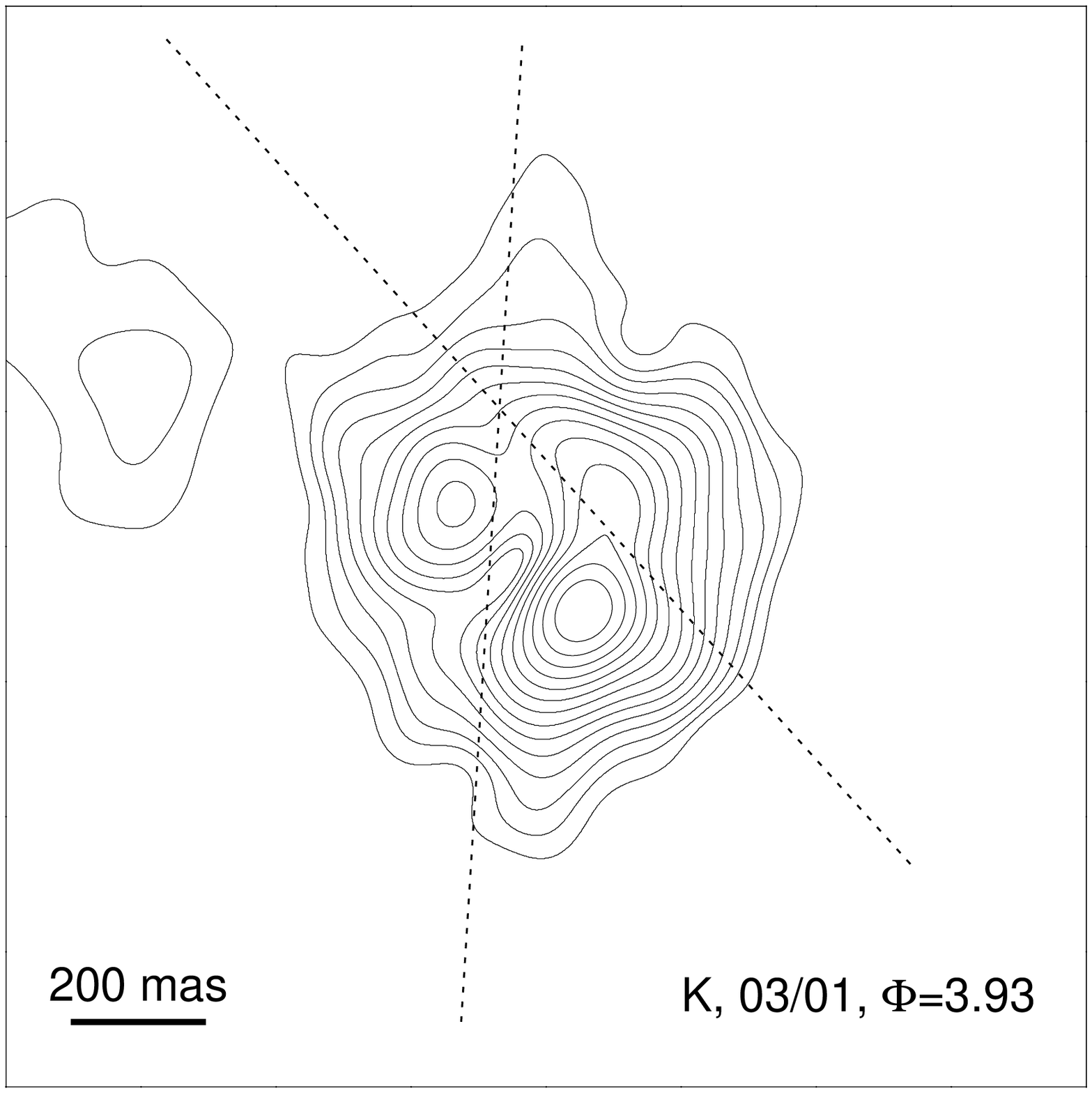}}
\end{center}
\caption{
{\it Left column:} $J$-band speckle images of \object{IRC\,+10\,216}
in April 1996 (top) and March 2001 (bottom)
reconstructed with the same resolution of 149 mas.
{\it Middle column:} $H$-band speckle images of \object{IRC\,+10\,216}
in January 1997 (top) and March 2001 (bottom)
reconstructed with the same resolution of 73 mas.
{\it Right column:} $K$-band speckle images of \object{IRC\,+10\,216}
in January 1997 (top) and March 2001 (bottom)
reconstructed with the same resolution of 90 mas.
Contour lines are shown from 0.3\,mag to 4.2\,mag
relative to the peak brightness
in steps of 0.3\,mag.
North is up and east is to the left.
The dashed lines indicate the biconical geometry of the cavities
according to the 2D radiative transfer model of Paper III; the lines intersect
at the position of component B.
}
\label{FJHKcont}
\end{figure*}

New bispectrum speckle-interferometry observations of \object{IRC\,+10\,216}
were carried out with the Russian 6\,m telescope at the Special Astrophysical
Observatory in September 1999, October 2000, and March 2001.
The speckle interferograms were
recorded with our HAWAII speckle camera
in the $J$, $H$, and $K$ bands.
The  speckle transfer function was derived from
speckle interferograms of unresolved reference stars.
Table~\ref{obstab} lists the observational parameters of these measurements
(1999--2001) complemented with the data of Paper II (1995--1998).
Images of \object{IRC\,+10\,216} with resolutions of
50\,mas ($J$),  56\,mas ($H$), and 73\,mas ($K$)
were reconstructed from
the speckle interferograms using the  bispectrum speckle-interferometry method
(Weigelt \cite{Wei77}, Lohmann et al.\ \cite{LohWeiWir83},
Hofmann \& Weigelt \cite{HofWei86}).
The modulus of the object Fourier transform
(visibility function) was determined  with the speckle interferometry method
(Labeyrie \cite{Lab70}).
The $H$- and $K$-band images are diffraction-limited,
the resolution in $J$  is only slightly (14\%) below the
diffraction limit.
Together with the data
of Paper II, a time
series of 8 $K$-band images, taken between 1995 and 2001, is
now available. Additionally, two epochs are covered in $J$ (1996, 2001) and $H$ (1997, 2001).
Figures~\ref{FKima}--\ref{FJHKcont} show
the reconstructed $K$-band images of the innermost
region of  \object{IRC +10 216} for all 8 epochs between
1995 and 2001, the most recent $J$, $H$, and $K$ images of March 2001,
and a comparison of the $J$, $H$, and $K$ reconstructions
from 1996/97 and 2001.
\subsection{Evolution in the $K$ band from 1995 to 2001} \label{Sobskevol}
The $K$--band images (Fig.~\ref{FKima}) show
several compact components within a 200 mas radius. 
We denote these components as A, B, C, and D in the order of decreasing
brightness in the 1996 image (cf.\ Paper I).
As already shown in Paper II,
the dust shell is dynamically evolving
and changes its shape on a timescale of $\sim 1$\,yr.
For instance, the apparent separation of the two
initially brightest components A and B increased
from $\sim 191$ mas in 1995 to $\sim 351$ mas in 2001.
At the same time, B has faded and almost disappeared in 2000
whereas the initially faint components C and D have become brighter.
In 2001, the intensity level of C increased to almost 40\%
of the peak intensity of A. 
The components A and C appear to have started merging in 2000.

During these six years, the compact and bright core of  component A
has been changing its shape continuously.
While it was almost spherically symmetric
in 1995, it is getting more elongated in the following years.
For example, the ratio of minor to major axis
drops to  $\sim 0.77$ during the years 1998 and 1999,
with the major axis being aligned along the
position angle $-48\degr$ (measured at 35\% intensity level).
The elongation of the inner core of A stays roughly constant.
Additionally, the morphology of the image in 2001 gives evidence of the
potential development of a bulge-like structure at PA=$-123\degr$ relative to
the center of A, which may
indicate the birth of a new component.

\begin{figure}
\epsfxsize=8,8cm
\mbox{\epsffile[56 12 549 710]{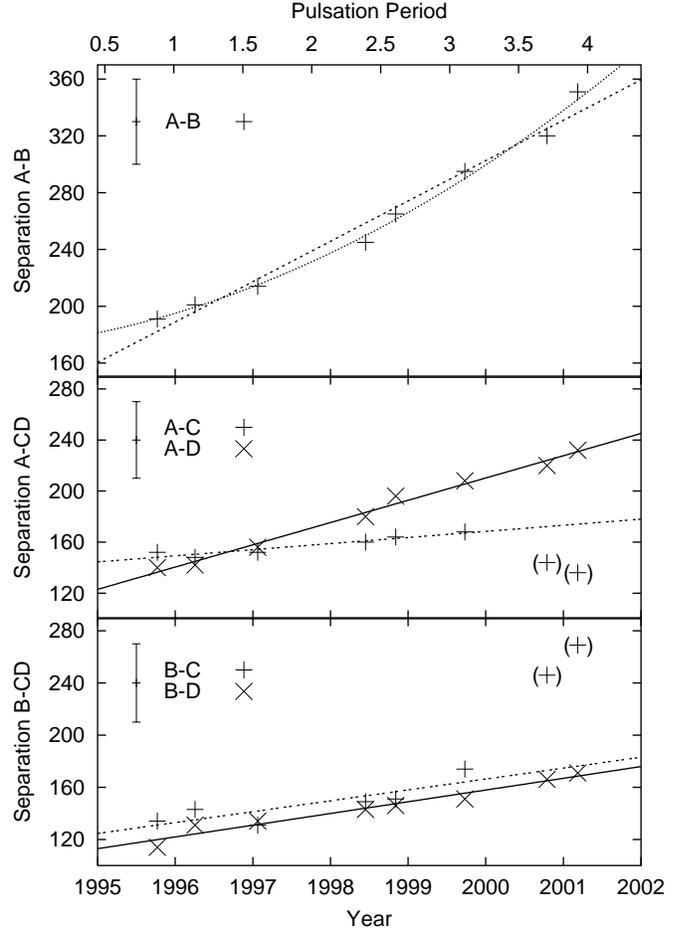}}
\caption{
Apparent separation of the components A, B, C, and D with respect to A and B
as function of time and photometric phase, resp.
The lines give least-squares fits of the
1995--2001 data resulting in average velocities of 28 mas/yr (A-B),
5\,mas/yr (A-C, for 1995--1999), 18 mas/yr (A-D),
8\,mas/yr (B-C, for 1995--1999), and 9 mas/yr (B-D).
The statistical velocity error bars amount to
$\pm 1$\,mas/yr (A-C, A-D, B-D), $\pm 2$\,mas/yr (A-B), and
$\pm 3$\,mas/yr (B-C).
The separations A-C and B-C in 2000 and 2001
(shown in parentheses) are not well-defined
due to the merging of A and C.
Also shown is a parabolic fit of the apparent increase of A--B
corresponding to an acceleration of $\sim$ 5\,mas/yr$^{2}$.
Note, however, that the existence of such acceleration
terms is not very certain
judged from the present data. The error bar given by the
pixel size (cf.\ Table~\ref{obstab}) is shown in the upper left corner
of each panel.
} \label{FsepABCD}
\end{figure}

The apparent separations of the various components with respect to A
and B are shown in Figure~\ref{FsepABCD}
as a function of time. 
The separations refer to the intensity maxima of the respective components
which evolve along approximately constant position angles in the 1990s.
Later, in 2000 and 2001, the position angles change moderately.

During the first 5 epochs   
the apparent separation between A and B increased by 23 mas/yr
(Paper II).
Interpreting this increase as real motion would lead to 14 km/s
within the plane of the sky (for a distance of  130\,pc). The corresponding
least-squares fit of the 1995--2001 data gives 28 mas/yr (18 km/s).
After 1999  component B became very faint and almost disappeared in 2001.
Therefore, its position in 2000 and 2001 is subject to larger
uncertainties.  
It should be noted that
the complete series from 1995 to 2001 can be best matched by a parabolic fit.
This may suggest an apparent acceleration
of $\sim$ 5\,mas/yr$^{2}$.
The fit is shown in  Fig.~\ref{FsepABCD}.
A similar acceleration term of  $\sim$ 3\,mas/yr$^{2}$ was inferred
by Tuthill et al.\ (\cite{TutEtal00}) from Keck telescope
$K$-band data for 7 epochs from 1997 to 1999.
%
%
%
%
%
The positions of
component D with respect to A are of particular interest since
they are well defined for the complete data set:
Whereas B considerably fades and
C appears to start merging with the bright core,
component D brightens and is well separated from A
in all images.
In Paper II,
we derived a velocity of
21\,mas/yr (13 km/s) for the
increase of the apparent separation A-D during 1995--1998.
With the new data we obtain 
18\,mas/yr (11 km/s) for 1995--2001, as shown in  Fig.~\ref{FsepABCD}.
For the evolution of these components,
there is no evidence of acceleration.
The distance A-C
has only slightly changed during the 6 years. 
A linear regression yields 5\,mas/yr for 1995--1999.
After 1999, A and C started merging and the separations do not
follow the above relation anymore (see Fig.~\ref{FsepABCD}).

In our radiative transfer modeling
(Paper III), component A was identified
with the southern cavity of a bipolar structure, whereas the location
of the central star is at component B. In the
lower panel of  Fig.~\ref{FsepABCD} we therefore present the
apparent separations of the
components C and D with respect to component B.  Least-squares fits give
8 mas/yr (5 km/s, for 1995--1999)
for the separation increase of B-C and 9 mas/yr (6 km/s) for that of B-D.
Due to the merging of A and C, the separation B-C increases
strongly in 2000 and 2001. The position of B becomes ill-defined
in the later epochs due to its dimming.
%
%
%
\subsection{\hbox{Changes of the $JHK$\,images from\,1996/97\,to\,2001}} \label{Shjk}
Figure~\ref{FJHKima} shows the
$J$, $H$, and $K$ images of \object{IRC\,+10\,216} obtained in March 2001.
The resolution in $H$ and $K$ is diffraction-limited
(56 and 73\,mas, respectively) whereas
in $J$  86\% of the diffraction limit is reached (50\,mas).
While the brightness of the inner dust shell components (yellow regions) increases
with increasing wavelength, the brightness of the faint surrounding nebula
(red regions) stays roughly constant. In the $J$ image, faint nebular
structure on sub-arcsecond scales can be seen.
The north-eastern
arm (PA $\sim 51\degr$ with respect to component A)
appears to have a clumpy structure.
Various components can be identified in this structure
at different separations from the center. The innermost clump 
is located at a separation of 256 mas from the center of A,
almost coinciding with the position
of component D in the $H$ and $K$ images. A closer inspection reveals that
this clump, in turn, resolves into two sub-structures 
of similar size.
Beyond this double knot, 
two further clumps 
can be identified
at a separation of 378 mas and 522 mas, resp.
The inner bright core A appears to show a multi-component structure in the
$J$ band as well.
It consists of two components
separated by $\sim 60$\,mas at PA $\sim$ 150\degr.

Figure~\ref{FJHKcont} shows the $J$-, $H$-, and $K$-band
images of \object{IRC\,+10\,216}
in 1996/97 compared to those in 2001.
For a given band, the images are reconstructed with the same resolution
(149\,mas in $J$,
73\,mas in $H$, and 90\,mas in $K$).
The dust-shell morphology and evolution observed in $H$
are very similar to those in $K$.
For instance, the apparent separation A-D increased
from 163 mas in 1997 to 245 mas in 2001, at the same time
component A became elongated (axis ratio of 0.65 in 2001) and a bridge
between A and C has formed.
In the $J$ band,
where one predominantly
observes scattered light,
the images changed significantly as well.
As in the other bands, component A became
elongated along the same position angle.
Three arms can be identified in north-western, north-eastern and south-western
direction (see also Fig.~\ref{FJHKima}).
The appearance of the northern arms becomes much more prominent between the
two epochs.
%
%

\begin{figure*}
\begin{center}
\epsfxsize=55mm
\mbox{\epsffile{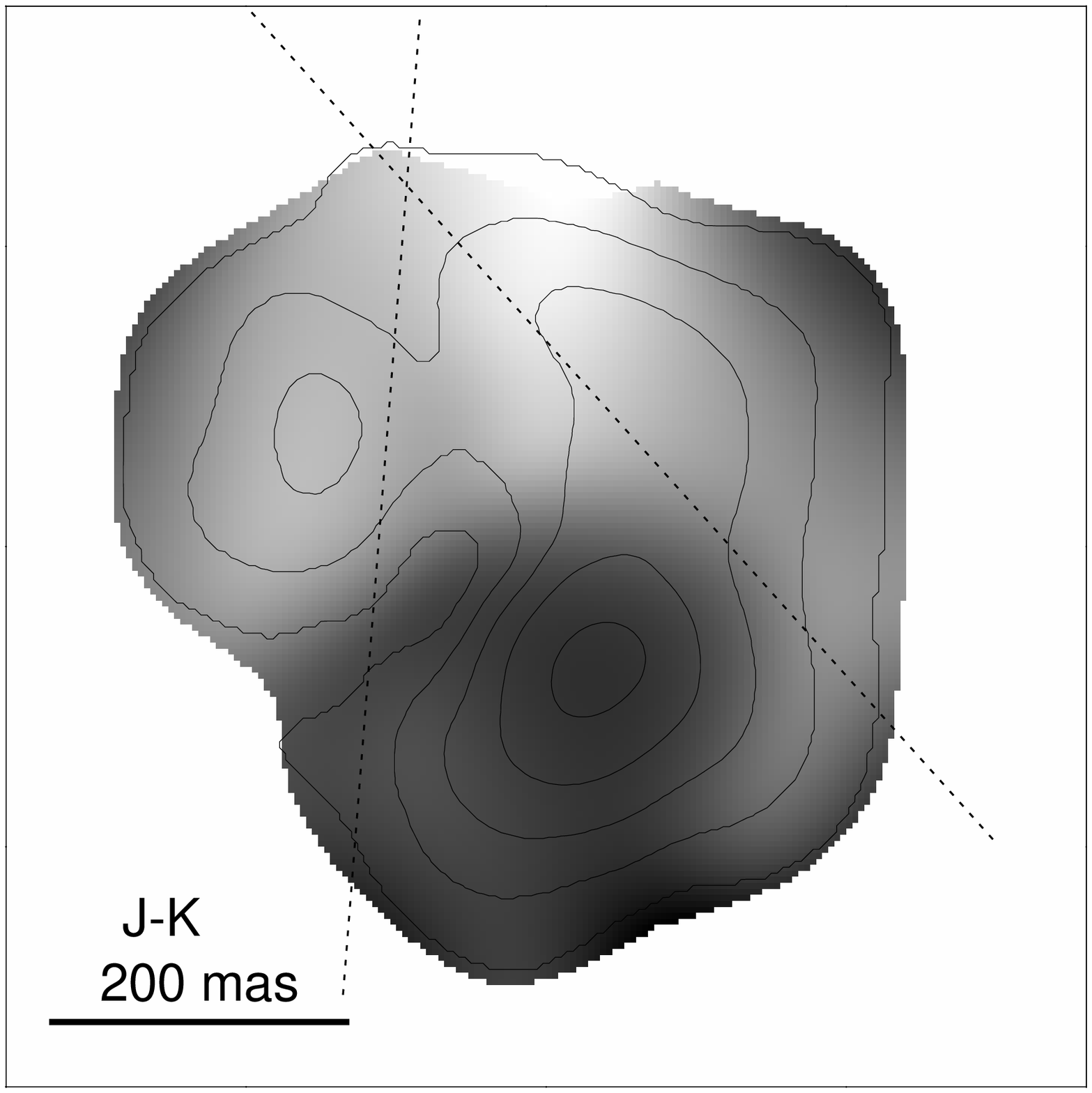}}
\epsfxsize=55mm
\mbox{\epsffile{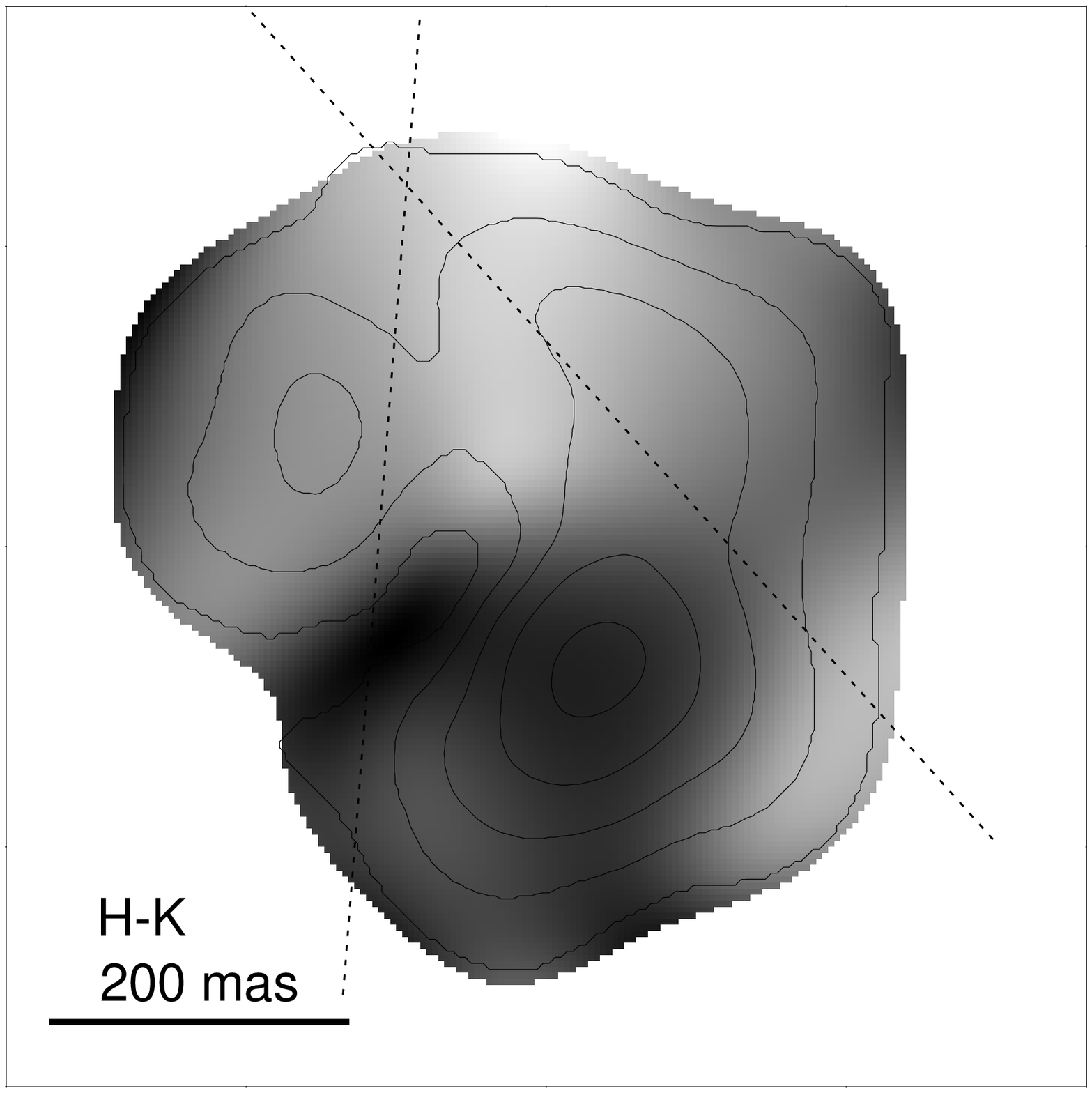}}
\end{center}
\caption{
$J{-}K$ (left) and $H{-}K$ (right)
color images (greyscale images)
of the central region of \object{IRC\,+10\,216}
in March 2001,  
confined to regions which are brighter than
2\% of the peak intensity in the $J$ and $K$ images
($J{-}K$ color image)
and 3\% of the peak intensity in the $H$ and $K$ images
($H{-}K$ color image), resp.
The resolution is 115 mas.
Contour lines refer to the $K$-band image and are shown from 0.2\,mag to 3\,mag
relative to the peak brightness
in steps of 0.7\,mag.
To construct the respective
color images, the $J$, $H$, and $K$ reconstructions were centered at the
positions of maximum intensity.
White regions refer to  magnitude differences of
7.4 mag ($J{-}K$) and 3.9 mag ($H{-}K$);
the regions shown in the darkest grey correspond to  magnitude differences of
5.0 mag ($J{-}K$) and 2.7 mag ($H{-}K$).
North is up and east is to the left.
Dashed lines indicate the biconical geometry of the cavities
according to the 2D  radiative transfer model of Paper\,III.
} \label{Fbild}
\end{figure*}
\subsection{
$J{-}K$ and $H{-}K$ color images}\label{Sobscol}
To compute color images, the $J$, $H$, and $K$ images observed in 2001
were reconstructed with the same resolution of 115~mas.
In our previous study (Paper II) we found that at previous epochs, when
both components A and B were clearly visible, the peak separations
in the $J$, $H$, and $K$ images were the same within a few milli-arcseconds
(at the same epochs). Therefore, we obtained almost identical color images
independently of whether the images were centered on component A or B. The
color images are not very sensitive to relative shifts within some
milli-arcseconds.

The resulting 
$J{-}K$ and $H{-}K$ color images for the
epoch of March 2001 are shown in Fig.~\ref{Fbild}
together with the intensity contour lines in the $K$ band.
The color maps are confined to regions which are brighter than
2\% of the peak intensity in the $J$ and $K$ images ($J{-}K$)
and 3\% of the peak intensity in the $H$ and $K$ images ($H{-}K$), resp.
The absolute magnitudes were calibrated with coeval photometry kindly provided
by B.\ Yudin. The integrated magnitudes in $J$, $H$, and $K$
(aperture 10$^{\prime \prime}$) are 5.98 mag, 3.51 mag, and 0.52 mag, resp.
In  Fig.~\ref{Fbild},
white color refers to magnitude differences of
7.4 mag ($J{-}K$) and 3.9 mag ($H{-}K$);
the darkest grey refers to magnitude differences of
5.0 mag ($J{-}K$), and 2.7 mag ($H{-}K$).
As in 1997
(Paper II),  component A turned out
to be clearly the bluest part of the inner dust-shell region, as expected
for a cavity illuminated by scattered light. Note that the star is
obscured by an optically very thick dust shell (${\tau}_V \approx 40$, cf.
Paper III) and therefore it must be located within the red area of the color
maps.
%
%
\section{Discussion} \label{Sdis}
\subsection{Dust and winds} \label{Swind}
On larger scales, most AGB stars are surrounded by dust shells of
spherical symmetry. This holds also for the dust shell of
\object{IRC\,+10\,216}  (Mauron \& Huggins \cite{MauHug99,MauHug00}).
However,  on  sub-arcsecond scales this symmetry appears to break down for
giants in very advanced stages of their AGB evolution, with
\object{IRC\,+10\,216} being their most prominent representative.
The 2D radiative transfer modelling
presented in Paper III strongly suggested that
the central star of \object{IRC\,+10\,216} is surrounded by an optically
thick dust shell with bipolar cavities of a full opening angle of
$36^{\rm o}$ tilted with the southern lobe towards the observer (although
there may exist different interpretations, see
Sect.~\ref{Sintro}). The bright compact component A is not the direct light
from the central star but is the southern cavity of this bipolar structure
dominated by scattered light. According to this model, the carbon star is at
the position of the fainter northern component B.

This dust-shell model and the
possible existence of acceleration within the cavities may
be understood in terms of the beginning operation of interacting winds
which later lead to the shaping
of planetary nebulae (Kwok \cite{Kwok82}; Balick \cite{Bal87}).
Within this scenario a fast stellar wind interacts with the
fossil, slow AGB wind. The AGB mass-loss is thought to be aspherical
and to take preferentially place in the equatorial plane.
Accordingly, the polar regions show lower densities than the equatorial
regions. Later,   
after a fast wind has developed, the less dense
polar material is carved out leading to the formation of outflow cavities.
This fast wind can be expected to turn on slowly at the very end of the
AGB evolution resulting into the formation of
biconical dust-shell structures on sub-arcsecond scales
as observed for \object{IRC\,+10\,216}.
%

The high spatial and temporal resolution of the present observations reveals
details of the  mass-loss process even
in the immediate vicinity of the dust condensation zone.
The monitoring, covering more than  3 pulsation
periods, shows that the structural variations are not related to the
stellar pulsation cycle in a simple way.
This is consistent with the predictions of
hydrodynamical models that enhanced dust formation takes place on a timescale
of several pulsation periods (Fleischer et al.\ \cite{FleiEtal95}).
In Sect.~\ref{Sdust} simplified dust formation
models are compared with the  observations.
\subsection{Observed changes interpreted by 2D modeling} \label{S2dmod}

The large number of epochs available now seems to make the interpretation of
the sub-arcsecond structures and their evolution even more difficult and
puzzling than before. Here we suggest a physical picture of what presently is
going on in this object based on the results of our previous 2D radiative
transfer modeling (Paper~III). The model was designed to derive the structure
and physical properties of the star and its dusty envelope at a single moment,
corresponding to the third epoch (January 23, 1997) of our images. Our
simplified attempt to handle the difficult problem of self-consistent {\em
time-dependent} multidimensional radiative transfer modeling will be presented
elsewhere (Men'shchikov et al. 2002, submitted).

\subsubsection{The star and its evolving environment}

Key point to the understanding of the structure and evolution of the
sub-arcsecond environment of \object{IRC\,+10\,216} is the knowledge of the
position of the central star in the images. As we have suggested in the
previous modeling (Paper~III), the brightest compact peak A seen in all images
is not the direct light from the central star. The star is most likely
located at the position of the fainter peak B, whereas component A is the
radiation emitted and scattered in the optically thinner conical cavity of the
optically thick bipolar dust shell. The even fainter components C and D in the
$H$ and $K$ images were identified with smaller-scale deviations of the density
distribution of the circumstellar environment from axial symmetry
(Fig.~\ref{FKima}; Fig.~2 in Paper~III). These inhomogeneities are less opaque
than other, more regular regions of the compact dense shell.

An alternative interpretation of the morphology of our images of
\object{IRC\,+10\,216} would be that components A and B are the lobes of a
bipolar nebula and that the star is located in the dark region {\em between}
them. However, this interpretation disagrees with our radiative transfer
modeling presented in Paper~III. A reason for this is that the optically
thick dust which would form the bipolar lobes cannot exist so close to the
star, at {\em half} the distance between the components. Thus, the star is
probably located either at A or B. The fan-shaped morphology of component A in
the $J$- and $H$-band images, as well as other evidence presented in Papers II
and III, strongly suggest that the star is located at the position of B.

Having defined the stellar position at component B, we can now interpret the
evolving appearance of \object{IRC\,+10\,216} in terms of an increased
mass-loss rate during the last $\sim$ 30 years. One of the main features of the
entire 8-epoch sequence of imaging is the relative fading of the stellar
component B (Figs.~\ref{FKima}, \ref{FJHKima}). Other changes seen in the
images are the increasing distances of components A, C, D from the star
(Fig.~\ref{FsepABCD}) and their greatly varying shapes. Our radiative transfer
modeling (Paper~III) has shown that the observed components are {\em cavities}
in the dense opaque shell; therefore, the observed motion requires an
additional discussion (Sect.~\ref{Motion}).

\subsubsection{The near-IR images and model cavities}

In order to better visualize the relative location of the components, we have
shown the structure (conical cavities) of our model (Paper~III) in the observed
images by dashed lines in Figs.~\ref{FJHKcont} and \ref{Fbild}. The lines
intersecting at the position of the central star and making an angle of
46{\degr} between them, outline the biconical geometry of the cavities.

The April 1996 $J$-band image in Fig.~\ref{FJHKcont} best exhibits the bipolar
geometry of \object{IRC\,+10\,216}. Clearly visible at this wavelength are the
fan shape of the brightest component A and even the scattered light from the
opposite (northern) cavity on the far side of the dense shell. Faint
direct stellar light is visible near the origin of the conical cavities.
The star (component B) appears much brighter in the January 1997 $H$-band
image (Fig.~\ref{FJHKcont}), whereas the fan shape of the cavity
(component A) is less prominent but still well visible at this
wavelength. The $H$ and $K$ images in 1997 agree very well with our model
presented in Paper~III. The March 2001 $H$ and $K$ images are qualitatively
similar to the older images obtained in 1997, but distorted by the appearance
of several fainter structures close to component D and by merging of the bright
component A with component C. The faint direct stellar light has become more
difficult to identify in $H$ and it has been buried in the enhanced dust
emission from all the other components in the March 2001 $K$-band image.

The most recent $J$ image of \object{IRC\,+10\,216} with a 3 times higher
resolution (50 mas, Fig.~\ref{FJHKima}) shows in much greater detail the fine
structure of the envelope. The image reveals a rather isolated, faint peak at
the position of the central star (marked in Fig.~\ref{FJHKima}). The following
evidence suggests that this peak is indeed the direct light from the star,
not just a clump that happened to be there: (1) its position angle relative to
A (PA $\approx 20$\,{\degr}) is the same as in all the images where the star
(component B) is clearly visible, (2) the $H$ image at the same epoch shows a
(less isolated) peak at the same position, (3) the distance of approximately
347 mas between the peak and the component A is consistent with the fit of
Fig.~\ref{FsepABCD}.

One might think that the faint stellar peak should be much better visible in
$K$ band (Fig.~\ref{FJHKima}), where optical depths are significantly lower
compared to $J$. However, the greatly enhanced hot dust emission in $K$ band
may well make it completely invisible. In fact, from the continuum of the model
stellar atmosphere used in Paper~III, we estimate that the stellar brightness
in $J$ and $K$ bands is approximately the same. The dust model of Paper~III
would predict that the star is a factor of $\sim 1.8 \times 10^3$ brighter in
$K$ than in $J$ band, whereas our new model of \object{IRC+10216} (Men'shchikov
et al., submitted), computed specifically for the latest epoch of March 2001,
predicts a larger factor of $\sim 1.3 \times 10^4$. On the other hand, from the
calibrated color images of Fig.~\ref{Fbild} we know that the nebula becomes
much brighter between $J$ and $K$ (and thus redder), by a factor of $\sim 1.4
\times 10^4$. Therefore, it is unlikely that in $K$ the stellar peak is
visible better than in $J$ band.

\subsubsection{Shapes of the near-IR color maps}

The high-resolution $J{-}H$, $J{-}K$, and $H{-}K$ color images in Fig.~\ref{Fbild}
observed in March 2001 confirm our interpretation that the bright components
A, C, and D are, respectively, the cavity and smaller-scale
inhomogeneities in the dense shell, {\em not} dense clumps of dust as it might
appear from the images alone (Figs.~\ref{FKima}--\ref{FJHKcont}). In fact, all
the bright components coincide with the blue areas in the color images, the
cavity A corresponding to the bluest spot. This is a natural
consequence of lower optical depths along those directions from the star, with
the ``hot'' stellar photons being scattered into the direction of the observer.
This is illustrated in Fig.~\ref{Fbild} by the dashed lines showing the
geometry of our model (Paper~III). The bluest, optically thinnest spot is
located precisely inside the conical cavity. The star, obscured by $\sim$
40 mag of visual circumstellar extinction, is situated in the red
area of the color images, as it was also in the previous epochs (Fig.~16 in
Paper~III).

\subsubsection{Moving dense layer or dust evaporation?}
\label{Motion}

The apparent motion of the components A, C, and D (Fig.~\ref{FKima}) could be
attributed either to the real radial expansion of the opaque dense layer with
several ``holes'' in the dense dust formation zone or to a displacement of the
dust formation radius due to evaporation of recently formed dust by a hotter
environment. Here, we analyze both processes and argue that the
temperature-induced displacement of the dust formation zone is acting in
\object{IRC\,+10\,216}. For simplicity, we consider a single dust formation
radius corresponding to the formation of carbon dust (cf. Paper~III).

One can explain the observed decreasing brightness of the star by assuming a
monotonically increasing mass-loss rate and, hence, higher densities and
optical depths of the wind in the dust formation zone. Higher mass loss and
continuing condensation of new dust in the wind out of the gas phase increase
the temperatures of the outflowing gas and dust due to backwarming. Increasing
temperatures affect the location of the inner dust boundary of the envelope via
dust sublimation, causing its displacement with a velocity that has nothing to
do with the outflow motion of the envelope.

Crucial to distinguishing between the real motion of a dense
layer and the temperature-induced shift of the dust formation zone are
estimates of the apparent velocities of the components A, C, and D relative to
the star. For the assumed distance of 130 pc, the linear fit in
Fig.~\ref{FsepABCD} gives for the brightest component A a velocity $v_{\rm A}
\approx 18$ km\,s$^{-1}$ in the plane of sky. On the basis of our model
(Paper~III), one can derive for the component a deprojected radial velocity
$v_{r{\rm A}} \approx 19$ km\,s$^{-1}$.

Since the deprojected radial velocity is higher than the observed (terminal)
wind outflow speed in \object{IRC\,+10\,216} of $v \approx 15$ km\,s$^{-1}$,
it is unlikely that the observed changes reflect just an expansion of a
dense layer in which grains are forming. In fact, the standard picture of a
stationary stellar wind predicts an acceleration of dust and gas within
distances by a factor of $\sim$ 2--5 larger than the dust formation radius
(Steffen et al.\ \cite{SteffEtal97}). Due to
the radiation pressure on dust grains, the wind velocity increases in this
transition zone, approaching asymptotically the terminal velocity at larger
distances. As our model (Paper~III) associates the observed components of
\object{IRC\,+10\,216} with the dust formation zone, we expect that dust and
gas are not yet fully accelerated there, i.e. that the radial outflow velocity
$v_r < 15$ km\,s$^{-1}$.

However, the deprojected velocity $v_{r{\rm A}} \approx 19$ km\,s$^{-1}$ is
significantly larger than the expected wind velocity in the dust formation
zone. Only if we assume an unlikely lower limit of 100\,pc (Becklin et al.
\cite{Becklin_etal1969}), does the deprojected velocity $v_{r{\rm A}}$ approach
(from above) $v \approx 15$ km\,s$^{-1}$, which is still too high for the dust
formation zone. If, however, the actual distance to \object{IRC\,+10\,216} is
larger than 130\,pc, one would obtain $v_r \ga 19$ km\,s$^{-1}$ and an even
larger discrepancy.

Moreover, there are reasons to believe that the acceleration depicted by the
parabolic fit in Fig.~\ref{FsepABCD} is real and that the actual apparent
motion of A is now as fast as $v_{\rm A} \approx 26$ km\,s$^{-1}$. The
corresponding deprojected velocity is then $v_{r{\rm A}} \approx 28$
km\,s$^{-1}$, much higher than the expected wind speed, for any realistic
distance to \object{IRC\,+10\,216}. Taken in context with increasing optical
depths in the shell, this suggests that the observed motions are caused
by the rapid dust evaporation due to backwarming and higher temperatures in the
dense environment formed by the increased mass loss.

We believe that a reasonable interpretation of the observed changes in
\object{IRC\,+10\,216} would be the following picture. During the recent
period of increasing mass-loss which started $\sim$ 20--30 years ago, a compact
dense shell has formed around the star. The mass-loss rate was recently as
high as $\dot{M} \approx 10^{-4} M_\odot$\,yr$^{-1}$ (Paper~III) and the
innermost regions of the dust shell are expanding outward at a local wind
velocity $v \la 10$ km\,s$^{-1}$. Dust formation continues in the expanding
material, thus increasing its optical depth and obscuring the central star.
The optical depths in the polar regions remain significantly smaller than in the
other regions of the dense shell, making the cavity (A) and the other
components relatively brighter than at previous epochs. As the dense,
increasingly optically thick dusty shell expands, steeply rising temperatures
inside it (due to the backwarming from the steepening density front) inhibit
further dust condensation and evaporate outflowing grains. In effect, these
processes have been shifting recently the dust formation radius outward with
an average velocity $v_r \approx 19$ km\,s$^{-1}$ (or as high as $\sim 30$
km\,s$^{-1}$ in 2001, if the apparent acceleration measured in this work is
real). One can predict that the star will remain obscured until $\dot{M}$
starts to drop back to lower values. In a few years from that moment, we could
probably be witnessing the star (B) reappearing whereas the cavities becoming
relatively fainter.
\subsection{Dust formation models and the fading of B} \label{Sdust}
If the star is located at the position of component B, as is suggested by
the two-dimensional radiative transfer model (see Sect.~\ref{S2dmod} and
Paper III),
\nocite{mbbow2001} the fading of B might be caused by the formation of new
dust along the line of sight towards the star.
To investigate, whether such a scenario would be
capable of explaining the observed time scale of the fading of B, we
constructed a very simple gas box model. The idea is to follow the
process of carbon grain formation in a gas element moving away from
the star at a constant velocity. To characterize the changing
thermodynamic conditions experienced by the gas element, we assume a
power-law gas temperature stratification

\begin{equation}
T^{\rm g}(r) = T^{\rm g}_{0} \left(\frac{R_0}{r}\right)^{\alpha}
\label{tempeq}
\end{equation}
and evaluate the gas density structure from mass conservation in a stationary,
spherically symmetric configuration
\begin{equation}
\rho^{\rm g}(r) = \frac{\dot{M}}{4 \pi v}\frac{1}{r^2} =
\rho^{\rm g}_{0} \left(\frac{R_{0}}{r}\right)^{2} \,.
\label{rhoeq}
\end{equation}
For the co-moving gas element, the time coordinate is given by
\begin{equation}
{\rm d} t = \frac{{\rm d} r}{v}\,\,\,.
\end{equation}

In this gas element, we consider heteromolecular formation and
growth of carbon grains which we compute according to the moment method
derived by Gail \& Sedlmayr~(1988)\nocite{gs88}.
The growth process includes reactions with the molecular species
C, C$_2$, C$_2$H, and C$_2$H$_2$.
In order to calculate the concentrations of the
relevant carbon-bearing molecules, chemical equilibrium in the gas phase is
assumed. We consider a carbon-rich element mixture with otherwise solar
abundances.

\begin{figure}
\begin{center}
\epsfxsize=8.8cm
\epsffile[18 148 445 664]{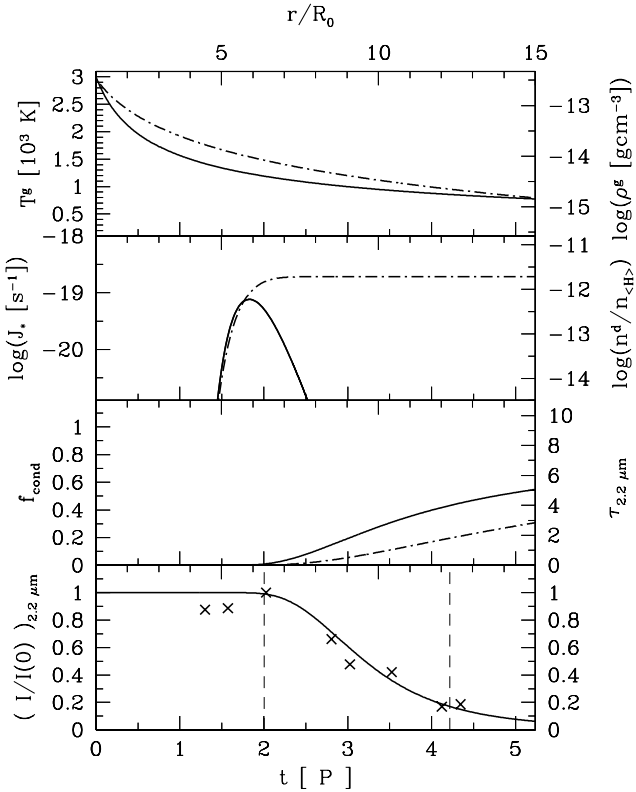}
\end{center}
\caption{Temporal evolution of the gas box.
The time coordinate is given in units
of $P=649$\,d, the pulsation period
of IRC~+10216.
{\it Upper panel}: gas temperature (Eq.~(\ref{tempeq}), solid line,
l.h.s. ordinate)
and density (Eq.~(\ref{rhoeq}), dash-dotted line, r.h.s. ordinate);
{\it Second panel}: dust nucleation rate $J_{*}$ (solid line, l.h.s. ordinate)
and number of dust grains per hydrogen nucleus
(dash-dotted line, r.h.s. ordinate);
{\it Third panel}: degree of condensation $f_{\rm cond}$
(solid line, l.h.s. ordinate) and optical depth
at 2.2\,$\mu$m (dash-dotted line, r.h.s. ordinate);
lower panel: normalized emergent intensity at 2.2\,$\mu$m (solid line) and
observed intensity of component B, normalized to its
maximum value (crosses). See text for more details.}
\label{gasboxplot}
\end{figure}

In Fig.~\ref{gasboxplot} the resulting temporal evolution of the gas box
is depicted for the following parameters:
$T_{0} = 3000\,$K at a radius $R_{0} = 3.15 \times 10^{13}\,$cm
(corresponding to
the Stefan-Boltzmann radius for $T_{0}$ and a stellar luminosity
of $15000\,L_{\odot}$), $\alpha = 0.5$ (i.e. an optically thin temperature
distribution),
$\dot{M} = 10^{-4}M_{\odot}$yr$^{-1}$ (reflecting a presently
increased mass-loss rate of IRC~+10216), a constant outflow velocity
of $v = 15\,$km\,s$^{-1}$, and a carbon abundance (by number)
of $\epsilon_{\rm C} = 1.5 \, \epsilon_{\rm O}$.

In the upper panel of Fig.~\ref{gasboxplot} the prescribed
time evolution of gas temperature (Eq.~(\ref{tempeq}))
and density (Eq.~(\ref{rhoeq})) in the gas box are shown.
The second panel depicts the nucleation rate $J_{*}$, i.e. the number of
seed particles formed per second and hydrogen
nucleus (solid line) and the number of dust grains per
hydrogen nucleus (dash-dotted line). In the third panel,
the degree of condensation, i.e. the fraction of condensible material that
is actually condensed is shown (solid line).
The dash-dotted line indicates the optical depth
at 2.2\,$\mu$m (measured from the star), that would build up in a stationary
situation. To calculate the dust extinction coefficient, we assume the small
particle limit of Mie theory. This is justified, since the mean particle
radius $\langle a \rangle$ reached
after 15\,$P$ is $5\times10^{-2}\mu$m, giving a maximum size
parameter $x = 2 \pi \langle a \rangle / \lambda = 0.14$ at the
considered wavelength of $\lambda = 2.2\,\mu$m. The optical properties of the
grains are described by the complex refractive index of amorphous carbon
tabulated in Preibisch et~al.~(1993)\nocite{poyh93}. In the lower panel we plot
the normalized intensity at 2.2\,$\mu$m that would emerge along the line of
sight from the increasingly obscured radiation source (solid line).
This intensity is calculated from the formal solution of the radiative
transfer equation:

\begin{equation}
\frac{I}{I_{0}} = e^{-\tau_{2.2\mu m}} +
\int_{0}^{\tau_{2.2\mu m}}\frac{B(T(t'))}{I_0}e^{-(\tau_{2.2\mu m} - t')}
\,{\rm d}t' \,\,\,.
\end{equation}

\noindent Here we assume thermal emission of the dust grains
($B$ is the Kirchhoff-Planck
function) and normalize the intensity to $I_{0} = B(T_{0})$.

For comparison, we also plotted in the lower panel of
Fig.~\ref{gasboxplot} the observed intensity of component B,
normalized to its
intensity in our 1997 speckle image (crosses). The intensity of component B
fades from its maximum value to 17\% of this value within about 1400\,d,
corresponding to 2.2 pulsation periods of IRC~+10216. This time
interval is indicated in the lower panel of Fig.~\ref{gasboxplot}
by the vertical dashed
lines.  The comparison shows that  dust condensation in front of the
star in fact can reproduce the observed time scale of the fading of
component B for realistic values of the parameters characterizing our
simple toy model.

In Table~\ref{timetab} the timescale $\tau$
of the intensity drop to 17\% of
the initial value is summarized for
various parameter combinations. We note a steep dependence of $\tau$
on the C/O ratio, i.e. on the amount of material available for dust
formation. Consistent model calculations for dust driven outflows
(e.g., Winters et~al.~2000b, Arndt et~al.~1997, Dominik et~al.~1990)
\nocite{wljhs2000,afs96,dgsw90a} show, that the outflow velocity of a
carbon-rich dust-driven wind also depends strongly on the C/O
ratio. For the observed outflow velocity of \object{IRC~+10216} of $\sim
15\,$km\,s$^{-1}$, C/O ratios in the range $(1.20 \la {\rm C/O} \la
1.60)$ are required. Therefore, Table~\ref{timetab} implies, that the
present day mass-loss rate of \object{IRC~+10216} in fact should have increased
considerably above the ``canonical'' value of (a few) $10^{-5}M_{\odot}$/yr
derived from CO rotational line observations
 probing the outer and therefore, older parts of the circumstellar shell
(e.g., Sch{\"o}ier \& Olofsson 2001)\nocite{so2001} or from CO infrared line
profiles observed about one decade ago (e.g., Winters et~al.~2000a).
\nocite{wkgs2000}

\begin{table}
\caption[]{Parameters of the gas box model and resulting fading time scales
for an assumed outflow velocity of $v=15\,$km\,s$^{-1}$}
\label{timetab}
\begin{flushleft}
\begin{center}
\begin{tabular}{ccccccc}
\hline\noalign{\smallskip}
$\alpha$ & $T_{0}$ & $\rho_{0}$ & $\dot{M}$    & C/O & $\tau$ \\
&   K   & $10^{-13}\,$g/cm$^{3}$ & $M_{\odot}/$yr & & 649\,d\\
\noalign{\smallskip}\hline\noalign{\smallskip}
0.5      &  2800 & 2.95 & $ 10^{-4}$ & 1.5 & 1.83\\
0.5      &  2800 & 2.95 & $ 10^{-4}$ & 1.4 & $>13$\\
0.5      &  2800 & 2.95 & $ 10^{-4}$ & 1.3 &$\rightarrow \infty$\\
\noalign{\smallskip}
0.5      &  2500 & 4.68 & $ 10^{-4}$ & 1.5 & 0.65\\
0.5      &  2500 & 4.68 & $ 10^{-4}$ & 1.4 & 1.79\\
0.5      &  2500 & 4.68 & $ 10^{-4}$ & 1.3 &$\rightarrow \infty$\\
\noalign{\smallskip}
0.5      &  2500 & 0.47 & $ 10^{-5}$ & 2.1 & 1.58\\
0.5      &  2500 & 0.47 & $ 10^{-5}$ & 2.0 & 3.05\\
0.5      &  2500 & 0.47 & $ 10^{-5}$ & 1.9 &$\rightarrow \infty$\\
\noalign{\smallskip}
0.5      &  3000 & 3.37 & $ 10^{-4}$ & 1.6 & 1.04\\
0.5      &  3000 & 3.37 & $ 10^{-4}$ & 1.5 & 2.22\\
0.5      &  3000 & 3.37 & $ 10^{-4}$ & 1.4 & $\rightarrow \infty$\\
\noalign{\smallskip}
0.4      &  3000 & 3.37 & $ 10^{-4}$ & 1.6 & 2.59\\
0.4      &  3000 & 3.37 & $ 10^{-4}$ & 1.5 & 7.53\\
\noalign{\smallskip}
0.6      &  3000 & 3.37 & $ 10^{-4}$ & 1.5  & 1.15\\
0.6      &  3000 & 3.37 & $ 10^{-4}$ & 1.4  & 5.01\\
\noalign{\smallskip}
0.7      &  3000 & 3.37 & $ 10^{-4}$ & 1.5 & 0.75\\
0.7      &  3000 & 3.37 & $ 10^{-4}$ & 1.4 & 2.54\\
\noalign{\smallskip}
0.7      &  3000 & 0.34 & $ 10^{-5}$ & 2.1 & 2.35\\
0.7      &  3000 & 0.34 & $ 10^{-5}$ & 1.9 & 9.62\\
0.7      &  3000 & 0.34 & $ 10^{-5}$ & 1.5 & $\rightarrow \infty$\\
\noalign{\smallskip}
\hline
\end{tabular}
\end{center}
\end{flushleft}
\end{table}

Once dust has formed in the circumstellar shell, its opacity would lead
to a pronounced backwarming and therefore, to a steeper temperature gradient
in the dust formation zone.
Accordingly, in our simple gas box model, lower temperatures would be
reached at the same density.
As a result, dust growth becomes more
efficient in this region and a reduced C/O ratio is sufficient to recover the
observed fading time scale (see Table~\ref{timetab}; compare the corresponding
entries for $\alpha = 0.5$ with $\alpha = 0.6$ and $0.7$).

A more realistic treatment of the dust formation and mass-loss process in
\object{IRC~+10216} would require the simultaneous
solution of the time dependent hydrodynamic equations, taking into
account the pulsation of the star, together with the equations
describing the dust formation process and the radiative transfer
problem. Such a consistent investigation will be presented in a forthcoming
paper.

An alternative interpretation of the fading of component B, together
with its increasing separation from A, could be based on a scenario,
where the star is located in the direction of component A, as argued
in previous studies by Weigelt et al. (\cite{WeiEtal98}), Haniff \& Buscher
(\cite{HanBus98}), and Tuthill et al. (\cite{TutEtal00}).  In this
case, component B would indicate thermal emission from a dust
cloud, ejected from the star in a direction almost perpendicular to
the line of sight. The temporal evolution of the
brightness of B would then be the result of a competition between the
initially increasing thermal emission due to ongoing dust formation in
the cloud and decreasing temperature of the cloud as it is moving away
from the star. Preliminary results for this scenario, which are based on
a consistent model calculation, are presented in
Winters et~al.~(2002)\nocite{wbhw2002}.
Although the time scale of the fading of component B is
reproduced quite well in this model, the calculated intensity ratio
between the expanding cloud and the star is smaller than the observed
intensity ratio between components B and A by about a factor of 3.

\section{Summary} \label{Scon}
We presented new
near-infrared ($JHK$) bispectrum speckle-interferometry monitoring
of the carbon star \object{IRC+10216} obtained during 1999--2001
with the SAO 6\,m telescope.
The  $J$-, $H$-, and $K$-band resolutions are
50\,mas, 56\,mas, and 73\,mas, respectively.
Together with the data of Paper II,
the available $K$-band observations cover now
8 epochs from 1995 to 2001
and show the dynamic evolution of the inner dust shell. Our images
show very good agreement with the images (1997--1999) reported by Tuthill
et al. (\cite{TutEtal00}).
Four main components within a 0\farcs2 radius can be identified,
which are surrounded by a fainter asymmetric nebula.
The apparent separation of the two
initially brightest components A and B increased
from $\sim 191$ mas in 1995 to $\sim 351$ mas in 2001.
At the same time, component B has been fading and almost disappeared in 2000
whereas the initially faint components C and D  became brighter.
There is weak evidence for an accelerated apparent motion of component B
with respect to A.
This might be related
to the beginning development of a fast polar wind or,
as favored by the radiative transfer calculations,
to rapid dust evaporation due to
backwarming effects.
%
%
The changes of the images can be related to
changes of the optical depth caused, for example, by
mass-loss variations or new dust condensation in the wind.
The observed relative motion of components A and B with a deprojected
velocity of 19 km\,s$^{-1}$ is most likely due to dust evaporation in
the optically thicker and hotter environment.

The present monitoring covers more than  3 pulsation
periods and shows that the structural variations are not related to the
stellar pulsation cycle in a simple way.
This is consistent with the predictions of
hydrodynamical models that enhanced dust formation takes place on a timescale
of several pulsation periods.
We have demonstrated, that formation of new dust along the line of
sight towards the star can explain the observed fading time
scale of component B, for reasonable values of the parameters
involved in our simple gas box model.
In particular, this dust formation calculation lends independent
support to the previous finding that the present-day mass loss rate of
\object{IRC~+10216} should be of the order of $10^{-4} M_{\odot}$/yr.
Further
high-resolution observations will be most important for testing different
views and models, and for better understanding of the evolution of this
complicated nebula.

\begin{acknowledgements}
The observations were made with  the SAO 6\,m telescope, operated by the
Special Astrophysical Observatory, Russia. We thank Boris Yudin for providing
near-infrared photometric data of \object{IRC\,+10\,216}.
This research has made use of the SIMBAD database, operated by CDS in
Strasbourg, and of NASA's Astrophysics Data System.
\end{acknowledgements}

\end{document}